\documentclass{aastex62}
\usepackage[flushleft]{threeparttable}
\usepackage{color}
\usepackage{float}
\usepackage[caption = false]{subfig}
\usepackage{gensymb}
\usepackage{graphicx}  
\usepackage{graphics} 
\usepackage{subfig}
\usepackage{stackengine}

\newcommand\aastex{AAS\TeX}

\submitjournal{ApJ}

\shorttitle{\aastex\ Stellar parameters of the first release of MaStar}
\shortauthors{Chen et al.}

\def\kms{\mbox{${\rm km}\,{\rm s}^{-1}$}}  

\begin{document}

\title{Stellar parameters for the First Release of the MaStar Library:  An Empirical Approach}

\correspondingauthor{Yan-Ping Chen}
\email{yc47@nyu.edu}

\author{Yan-Ping Chen}
\affiliation{New York University Abu Dhabi, P.O. Box 129188, Abu Dhabi, United Arab Emirates}
\affiliation{Center for Astro, Particle and Planetary Physics (CAP3), New York University Abu Dhabi, P.O. Box 129188, Abu Dhabi, UAE}
\author{Renbin Yan}
\affiliation{Department of Physics and Astronomy, University of Kentucky, 505 Rose St., Lexington, KY 40506-0057, USA}

\author{Claudia Maraston}
\affiliation{Institute of Cosmology \& Gravitation, University of Portsmouth, Dennis Sciama Building, Portsmouth, PO1 3FX, UK}

\author{Daniel Thomas}
\affiliation{Institute of Cosmology \& Gravitation, University of Portsmouth, Dennis Sciama Building, Portsmouth, PO1 3FX, UK}

\author{Guy S. Stringfellow}
\affiliation{Center for Astrophysics and Space Astronomy, Department of Astrophysical and Planetary Sciences, University of Colorado, 389 UCB, Boulder, CO 80309-0389, USA}

\author{Dmitry Bizyaev}
\affiliation{Apache Point Observatory and New Mexico State University, P.O. Box 59, Sunspot, NM 88349, USA}
\affiliation{Sternberg Astronomical Institute, Moscow State University, Universitetskij pr. 13, Moscow, Russia}

\author{Joseph D Gelfand}
\affil{New York University Abu Dhabi, P.O. Box 129188,  Abu Dhabi, United Arab Emirates}
\affil{Center for Astro, Particle and Planetary Physics (CAP3), New York University Abu Dhabi, P.O. Box 129188, Abu Dhabi, UAE}

\author{Timothy C. Beers}
\affiliation{Department of Physics and JINA Center for the Evolution of the Elements, University of Notre Dame, Notre Dame, IN 46556, USA}

\author{{Jos{\'e} G. Fern{\'a}ndez-Trincado}}
\affiliation{Instituto de Astronom\'ia y Ciencias Planetarias, Universidad de Atacama, Copayapu 485, Copiap\'o, Chile}

\author{Daniel Lazarz}
\affiliation{Department of Physics and Astronomy, University of Kentucky, 505 Rose St., Lexington, KY 40506-0057, USA}

\author{Lewis Hill}
\affiliation{Institute of Cosmology \& Gravitation, University of Portsmouth, Dennis Sciama Building, Portsmouth, PO1 3FX, UK}

\author{Niv Drory}
\affiliation{McDonald Observatory, The University of Texas at Austin, 1 University Station, Austin, TX 78712, USA}

\author{Keivan G. Stassun}
\affiliation{Department of Physics and Astronomy, Vanderbilt University, VU Station 1807, Nashville, TN 37235, USA}






\begin{abstract}

We report the stellar atmospheric parameters for 7503 spectra contained in the first release of the MaNGA stellar library (MaStar) in SDSS DR15.
 The first release of MaStar contains 8646 spectra measured from 3321 unique stars, 
 each covering the wavelength range 
3622 \AA\  to 10354 \AA\ with a resolving power of $R \sim$ 1800.    
In this work, we first determined the basic stellar parameters: effective temperature ($\rm T_{eff}$), surface gravity ($\log g$), and metallicity
($\rm[Fe/H]$), 
which best fit the data using an empirical interpolator based on the Medium-resolution Isaac Newton Telescope library of 
empirical spectra (MILES), 
as implemented by the University of Lyon
Spectroscopic analysis Software \citep[][ULySS]{Koleva08} package.
While we analyzed all 8646 spectra from the first release of MaStar, since MaStar has a wider parameter-space coverage than MILES, not all of these fits are robust. In addition, not all parameter regions covered by MILES yield robust results, likely due to the non-uniform coverage of the parameter space by MILES.
We tested the robustness of the method using the MILES spectra itself and identified a proxy based on the local density of the training set. With this proxy, we identified
7503 MaStar spectra with robust fitting results. They cover the range from 3179K to 20,517K in effective temperature ($\rm T_{eff}$), from 0.40 to 5.0 in surface gravity ($\log g$), and from $-$2.49 to $+$0.73 in metallicity ($\rm[Fe/H]$).
 
\end{abstract}




\section{Introduction} \label{sec:intro}

A stellar library is a collection of spectra of different stars, sharing the same wavelength range and resolution, that 
cover a certain parameter space of atmospheric properties. 
A stellar library can be theoretical (i.e., based on stellar atmospheric models) or empirical (based on spectral observations). 
Examples of theoretical stellar libraries include, but are not limited to,
\citet{Kurucz79} and its re-calibration by \citet{Lejeune98}, \citet{Zwitter04, Martins05, Munari05, Coelho05, Coelho07, Gustafsson08, Leitherer10, deLaverny12} and \citet{BOSZ}.
Examples of empirical libraries include 
Pickles \citep{Pickles85, Pickles98}, 
\citet{Diaz89}, \citet{Silva92}, Lick/IDS \citep{Worthey94}, \citet{Lancon2000}, STELIB \citep{stelibref}, ELODIE \citep{elodie}, INDO-US \citep{Valdes04}, CaT \citep{Cenarro01}, MILES \citep{milesref,Falcon-barroso11}, HST NGSL \citep{ngsl}, X-Shooter Stellar Library \citep[XSL,][]{XSL}, the NASA Infrared Telescope Facility (IRTF) Library \citep{Rayner09}, and the Extended IRTF library \citep{Villaume17}.

Stellar libraries play an essential role for a wide range of astrophysics applications. In particular, they serve 
as a reference for the classification and automatic analysis of large stellar spectroscopic surveys, and are 
fundamental ingredients for the models of stellar populations used to study the evolution of galaxies.

Initiated by the need to model the 
spectra of galaxies collected by the Mapping Nearby Galaxies at Apache Point Observatory survey  \citep[MaNGA,][]{Bundy15, Yan16}, 
which is one of the state-of-the-art spectroscopic surveys of SDSS-IV \citep{Blanton17}, 
we have developed the MaNGA stellar library (MaStar) to build a large, comprehensive stellar library that shares the 
same wavelength coverage of MaNGA galaxies and other SDSS spectra, i.e., 3622 \AA\  -- 10354 \AA\/, and
includes stars covering a wide range of stellar-parameter space with a resolving power of $R \sim 1800$.  We have piggybacked on the 
Apache Point Observatory Galaxy Evolution Experiment 2 (APOGEE-2) to reduce the cost of covering a large area of the sky, as well as 
many existing stellar spectroscopic surveys to preselect our targets according to their measured parameters
\citep[see][for details]{Yan19}. Briefly, we constructed the target list of MaStar from stars previously observed by several large stellar surveys, i.e.,
APOGEE-1, APOGEE-2 \citep{Majewski17}, SDSS/SEGUE \citep{yanny09}, and LAMOST \citep{cui12, deng12, zhao12}. 
For fields without a sufficient number of stars with existing stellar parameters from the literature, we also estimated
stellar parameters for millions of stars using PanSTARRS1  \citep{pansref} and 
APASS\footnote{\url{https://www.aavso.org/apass}}, as described in \citet{Yan19}. 

The MaStar project is obtaining stellar spectra that span a wider range of stellar-parameter space -- mass, effective temperature 
($\rm T_{eff}$), surface gravity ($\log g$), luminosity class, and chemical composition -- than any previously published empirical spectral library. Upon completion of the MaStar survey in 2020, at the end of SDSS-IV, the MaStar library will contain of order 10,000 unique stars. 
MaStar DR1 \citep{Yan19} is part of the SDSS DR15 and is publicly available at {https://www.sdss.org/dr15/mastar/}. 

Obtaining the stellar spectra is the first step in constructing a stellar library. The next step involves 
quantitatively deriving the fundamental stellar parameters from each spectrum. For applications that use individual spectra 
in the library, these parameters allow researchers to select appropriate stars for comparison. For stellar-population synthesis, 
these parameters are needed to associate each spectrum with a position on a stellar evolutionary track or isochrone. 
We have collected the literature parameters for $\sim 70\%$ of the 
stars in MaStar DR1. However, a large fraction, i.e., the remaining 30\% of MaStar targets do not have input stellar parameters.
Moreover, since the stars with known stellar parameters are collected from various stellar surveys, the mixing of different systematics may create in-homogeneity. We therefore aim to calculate a consistent set of stellar parameters
for all MaStar spectra. Furthermore, to ensure continuity with previous work on galaxy evolution \citep[e.g.,][]{Vazdekis10, Maraston11, Conroy12}, we require that our results are compatible with the MILES library, the current state-of-the-art in this field.
The parameters from this paper have been used for the first MaStar-based stellar-population 
models \citep[E-MaStar, where E stands for `empirical', to acknowledge the semi-empirical nature of the parameters determined here;][]{Maraston20}

In this work, we focus on determinations of 
three fundamental stellar parameters: effective temperature, $\rm T_{eff}$, surface gravity, $\log g$, 
and metallicity, $\rm[Fe/H]$.
In Section~\ref{secdata}, we briefly review the MaStar data. In Section~\ref{secmeth}, we present the method 
used to determine the stellar parameters for MaStar DR1. The stellar-parameter quality control is discussed in Section~\ref{secqc}.  
We describe the method used to validate our results in Section~\ref{secval}, and demonstrate consistency with literature parameters
in Section~\ref{secexqc}. We summarise our results and conclusions in Section~\ref{secconclusion}.\\

\section{data}\label{secdata}

The MaStar program was designed to acquire optical spectra using the MaNGA  fiber bundles in tandem with APOGEE-2N observations during APOGEE--led bright time observations.\footnote{MaStar has supplemented the main MaStar protocol with some MaStar-led plates. }
Therefore, the MaStar stellar spectra 
share the same instrument resolutions as the spectra obtained for MaNGA galaxies, 
making them ideal for constructing templates used to model the stellar continuum, and populations of these sources. 
Data reduction for MaStar spectra was 
performed by the MaNGA Data Reduction Pipeline \citep[DRP;][]{Law16}, an IDL-based software suite that produces final flux-calibrated data
cubes from the raw dispersed fiber spectra. We refer readers to \citet{Law16} for details about the DRP. After the data were reduced, we 
eliminated spectra with low signal-to-noise, bad sky subtraction, high scattered light, low point spread function (PSF)-covering fraction, uncertain radial velocity measurement, and/or those with problematic flux calibration. In addition, we also  visually inspected 
each spectrum using the Zooniverse Project Builder interface \footnote{\url{https://www.zooniverse.org/lab}}. A total of 28 volunteers from within the collaboration participated in order to check for issues of flux calibration, sky subtraction, telluric subtraction, emission lines, etc. The results are input to the DRP to assign the final quality flags.

\begin{figure}
\centering
\includegraphics[scale=0.60]{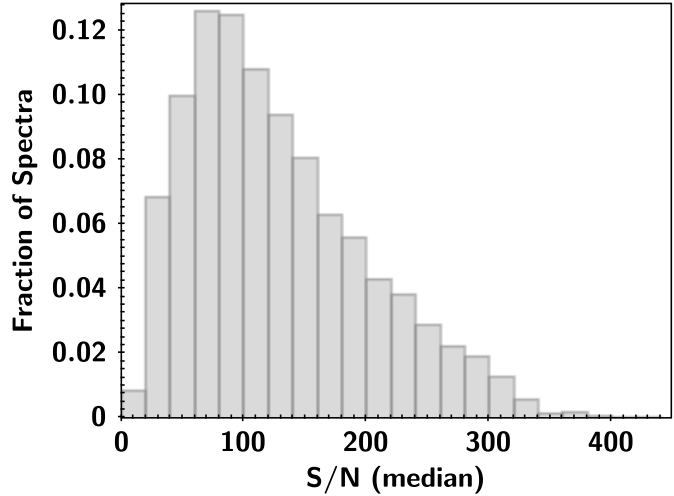}
\caption{Median signal-to-noise distribution of the 8646 MaStar spectra in DR1. Numbers are normalized by the total number of stars, so the bins represent fractions of the spectra.}
\label{fig:sndistri}
\end{figure}

The spectra we used in this work are from the first release of MaStar \citep{Yan19}. These spectra span a wavelength  
3622 \AA\  to 10354 \AA\ with a resolving power of $R \sim 1800$, and are provided in DR15 of SDSS-IV \citep{Aguado19}. For each spectrum, the calibrated flux, inverse variance of the flux, accurate line-spread-function (LSF), and mask are 
provided as a function of wavelength. In Figure~\ref{fig:sndistri}, we show the distribution of the median signal-to-noise (S/N) over the full wavelength coverage of all spectra contained in the first release of MaStar. Most ($\sim 90\%$) 
spectra have a mean S/N above 50. This release of MaStar contains 3321 unique stars with 
8646 spectra that have accurate flux calibration, within $\sim4\%$ \citep{Yan19}.

\section{Method}\label{secmeth}

 To determine the stellar parameters of the MaStar objects we use the full-spectrum-fitting package 
 ULySS\footnote{\url{http://ulyss.univ-lyon1.fr/}} \citep{Koleva08}. 
 This package performs $\chi ^2$ fitting between a template spectrum and the data, where the model spectrum 
is interpolated between a set of spectra with known stellar parameters of $\rm T_{eff}$, $\log g$, and $\rm[Fe/H]$.
The basic idea is that the interpolator approximates the normalized flux in each spectral bin as polynomials of $\rm T_{eff}$, 
$\rm [Fe/H]$, and $\log g$, with each set of polynomials defined for three groups of stars (OBA, FGK, and M).
Interpolation was applied in overlapping regions in between the groups in the parameter space. The algorithm allows the use of Legendre polynomials to 
account for residual extinction and/or flux--calibration systematics.

There are several interpolators published by the same group, e.g., the ELODIE interpolator \citep{Wu11}, the MILES
interpolator \citep{Prugniel11}, and the expanded MILES interpolator \citep{Sharma16}, 
which extends the MILES interpolator to include cool stars ($\rm T_{eff}
\leq$4500K) from the FEROS 
archive, with stellar parameters adopted from \citet{Allen73}, \citet{ Casagrande08}, and \citet{Neves13}. However, our results, based on the extended MILES interpolator 
show artificial wiggles in the cool dwarfs region. Therefore, we prefer to use an older version of the MILES 
interpolator \citep{Prugniel11}, 
which spans a wider range of stellar parameters 
 than the interpolator based on the ELODIE library.
 The MILES library also has a wider wavelength coverage than the ELODIE library. 
 The MILES stellar parameters from \citet{Cenarro07} were not used to build this interpolator, because that is based on a heterogeneous combination of 
 literature compilation and photometric calibration \citep{Prugniel11}. 
Instead, the MILES interpolator used in this work was built by recalculating the stellar parameters of MILES spectra 
using ELODIE 3.2 \citep{Wu11} as the reference for homogeneous stellar atmospheric parameters, with a  semi-empirical solution
\citep{Prugniel07} introduced to cover a better atmospheric-parameter range using theoretical spectra  
of hot stars with $\rm T_{eff} \geq 20,000K$ from \citet{Martins05}.
Some low-metallicity cool dwarfs from \citet{Coelho05} are included to widen the stellar-parameter coverage as well.

We estimate the stellar parameters by constructing an initial guess grid to locate the possible parameter range, i.e., 
$\rm T_{eff}$=[3500, 4000, 5600, 7000, 10000, 30000]K, 
$\rm [Fe/H]= [-1.7, -0.5, +0.5]$, and $\log g=$ [1.8, 3.8], so that the interpolator can find the closest parameter region to use as a starting point. 
Since the spectra of MaStar span the wavelength range of 3622--10354 \AA, we use the common wavelength range 
(i.e., 3622--7400\AA) between MILES and MaStar in the full-spectrum-fitting process. Error spectra, along with the data, were used to 
find the best-fit, with bad pixels masked. The initial fit is performed by finding the minimum $\chi^2$ between 
the observed spectrum and an interpolated MILES spectrum. We then allow the interpolator to search for a
 better solution in the parameter space within the grid by finding the final fit that has the best $\chi^2$ in the local solution space.

As mentioned by \citet{Prugniel11} and \citet{Wu11}, the line spread function (LSF) may affect the full-spectrum-fitting procedure.
We therefore calculate the difference between the LSF of each input MaStar spectrum and the intrinsic resolution. 
Several groups have calculated the intrinsic resolution of the MILES library following its publication.  
 \citet{Falcon-barroso11} derived FWHM=2.51\AA.  A similar value for the
 MILES resolution, FWHM =2.54\AA, was also calculated
 by \citet{Beifiori11}. Here we use FWHM=2.51\AA\ for the MILES templates. 
 The MILES interpolator is then convolved to the same resolution as the
 individual MaStar spectrum. 
As the LSF varies with wavelength, we accordingly convolved the MILES templates as a function of wavelength.
The best fit is obtained by comparing the MILES interpolated model spectra and the MaStar data with pixel
size 69.03 $\kms$, evenly spaced in logarithmic wavelength.  
A multiplicative polynomial is employed in the fitting process to match the overall shape between the 
data and the model. As mentioned in \citet{Yan19}, the MaStar spectra are not extinction corrected. 
Therefore, when performing the spectrum fitting, we need to assume a polynomial to 
correct for differences in the broadband shape due to extinction effects and/or flux-calibration residuals. 
We initially used a 10th-order multiplicative polynomial to account for these effects, and then reduce the  
order of the polynomial in future iterations to achieve better fits. The final polynomial is then multiplied 
with the data to match the assigned template. 
MaStar DR1 has $\sim 80\%$ of the stars with 
$E_{B-V} \leq 0.1$. However, this has no effect on the final parameters, as we rely on the relative ratios of the lines on the full spectrum fitting result.

\begin{figure*}[ht!]
\centering
\captionsetup[subfigure]{labelformat=empty}
  \subfloat[ ]{ \includegraphics[scale=0.45,angle=0]{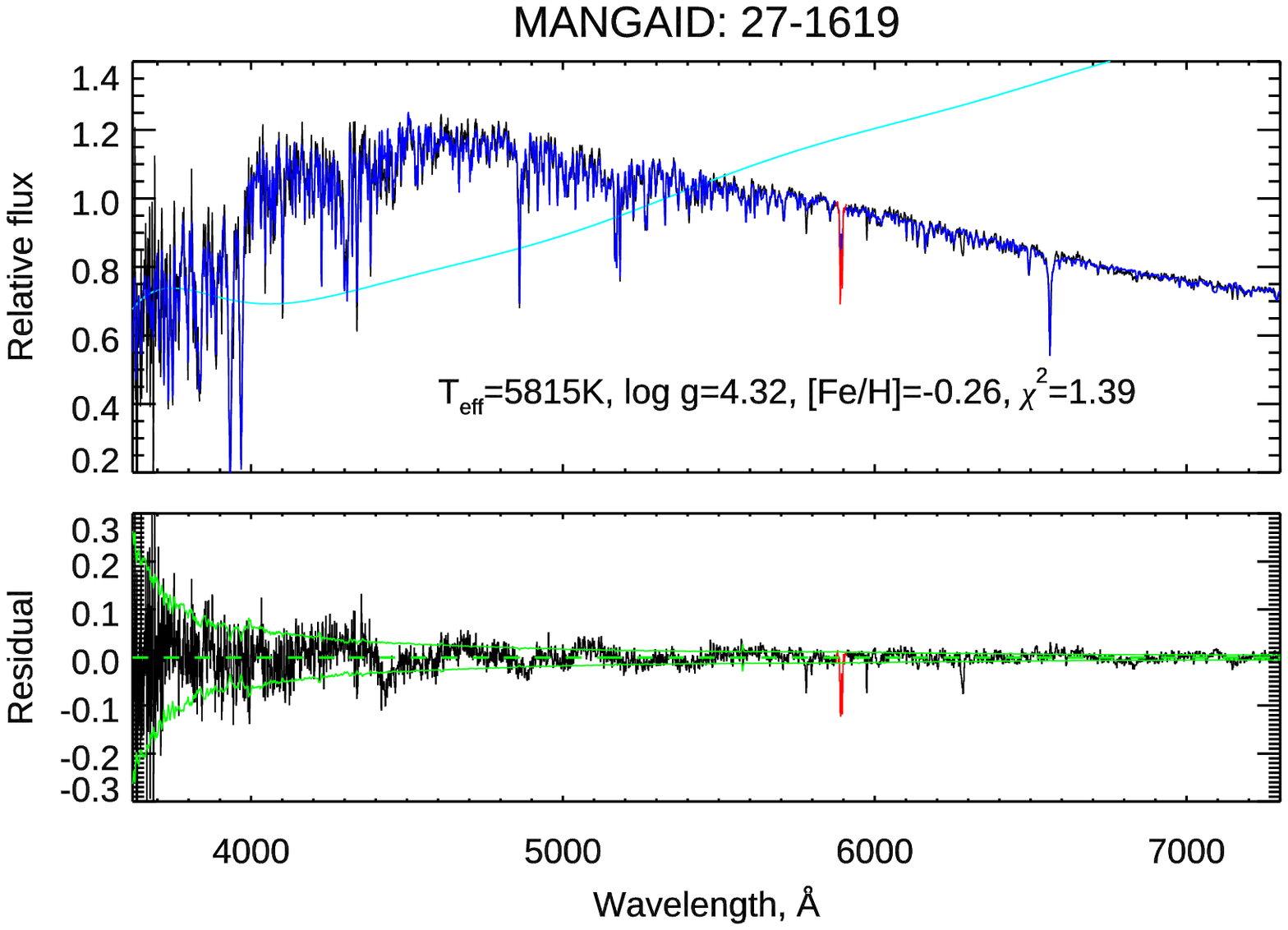}}
  \subfloat[ ]{ \includegraphics[scale=0.45,angle=0]{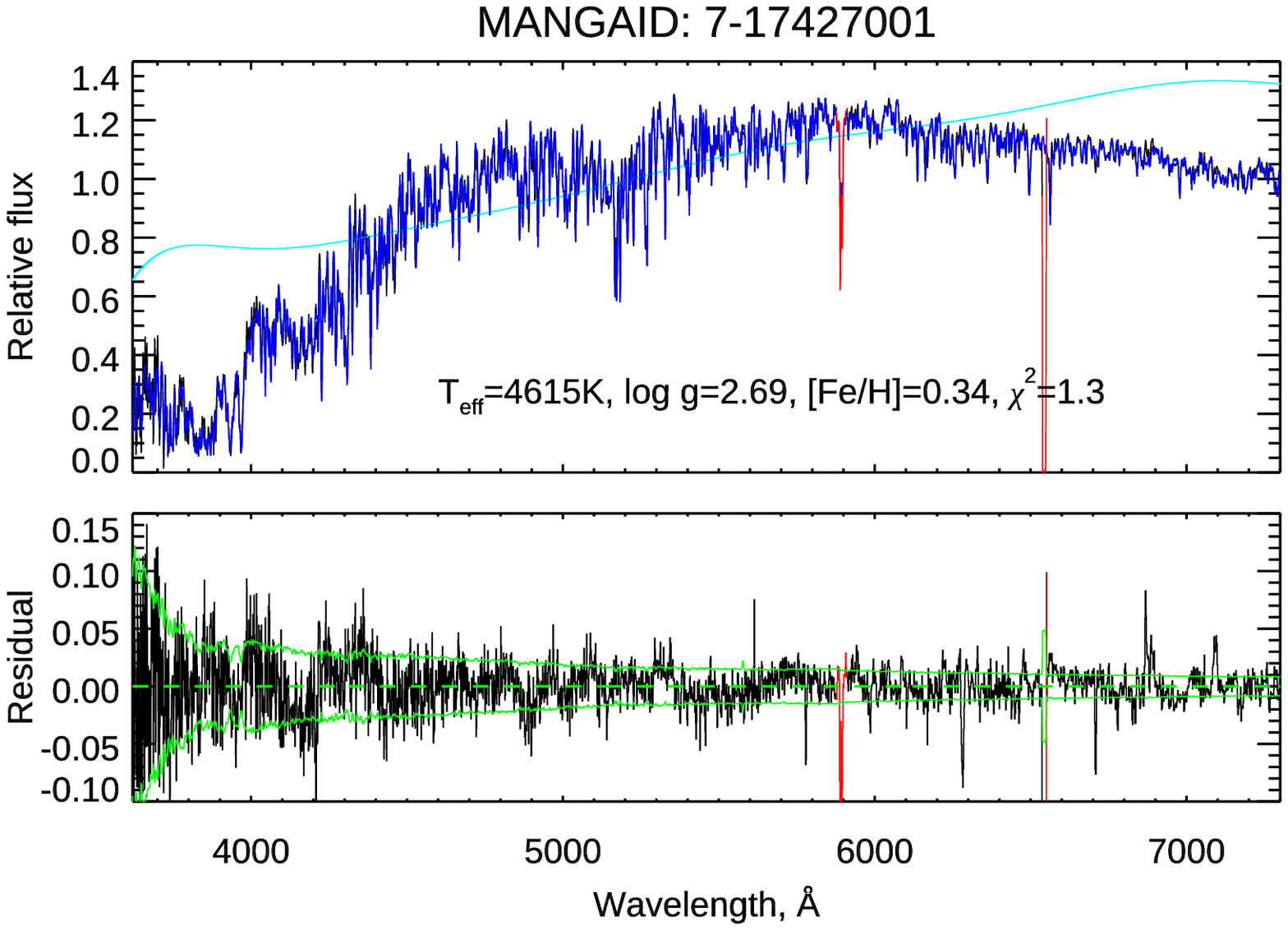}}\\
  \vspace{-2\baselineskip}
  \subfloat[ ]{ \includegraphics[scale=0.45,angle=0]{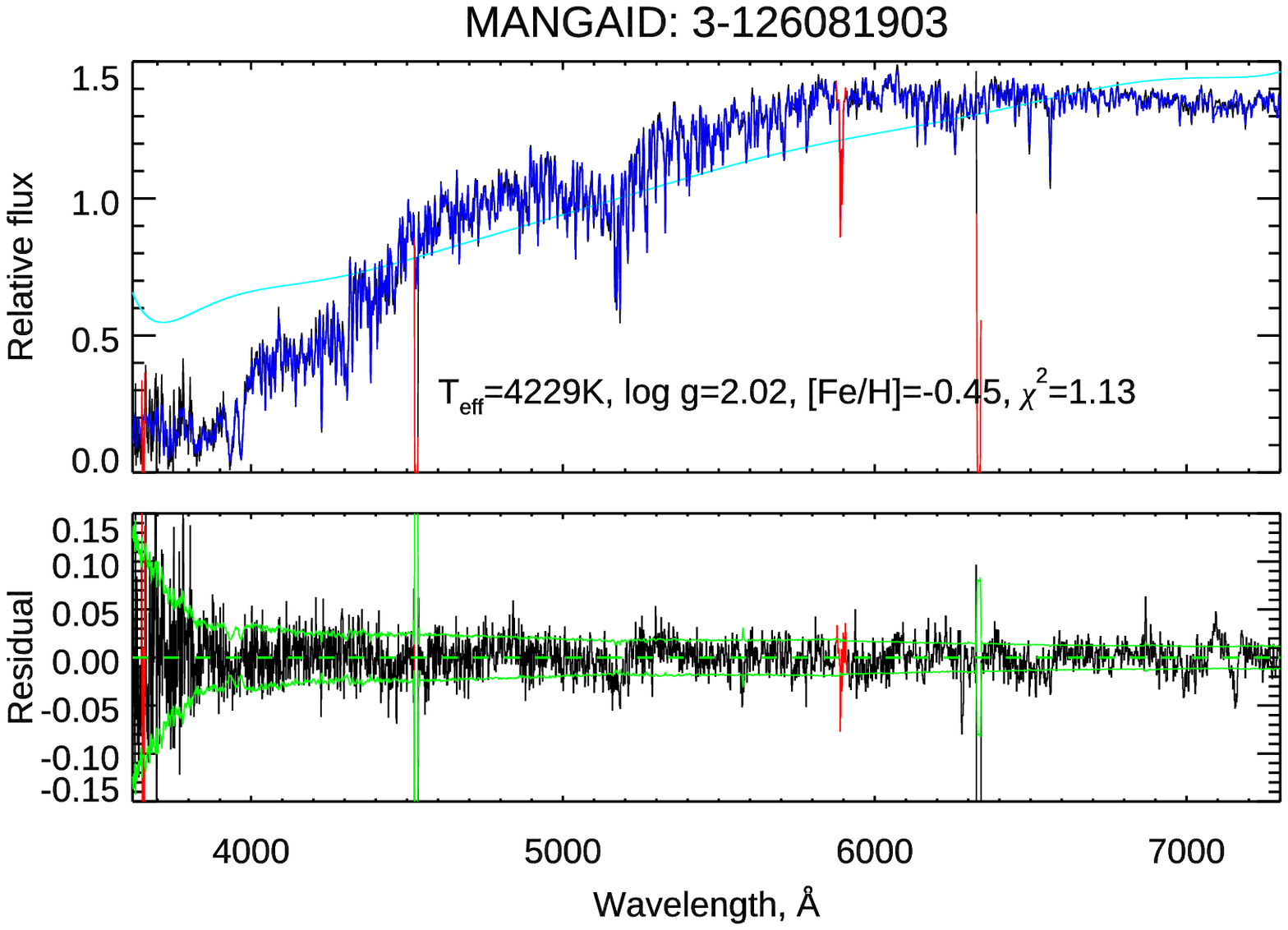}}
  \subfloat[ ]{ \includegraphics[scale=0.45,angle=0]{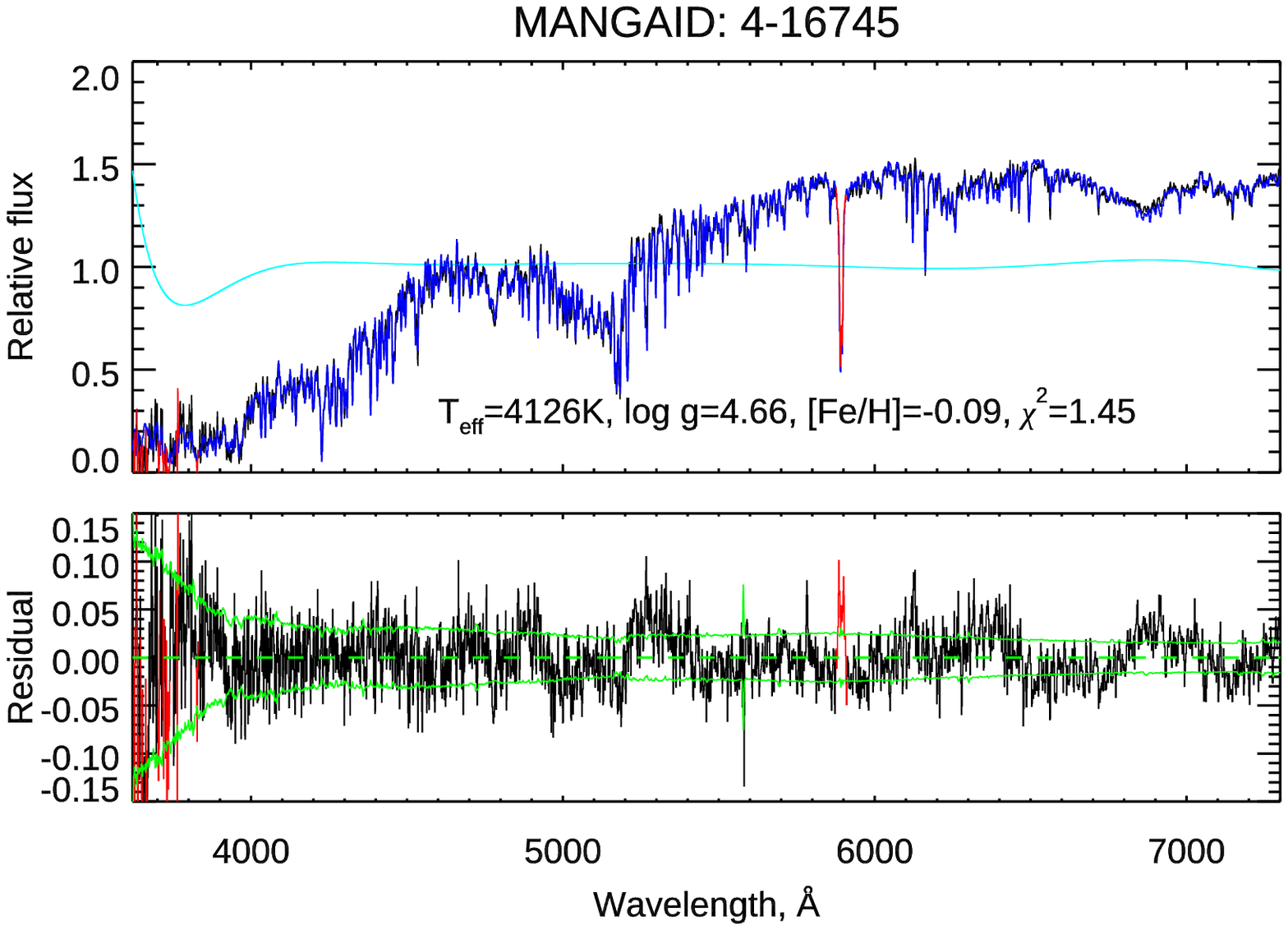}}\\
  \vspace{-2\baselineskip}
  \subfloat[ ]{ \includegraphics[scale=0.45,angle=0]{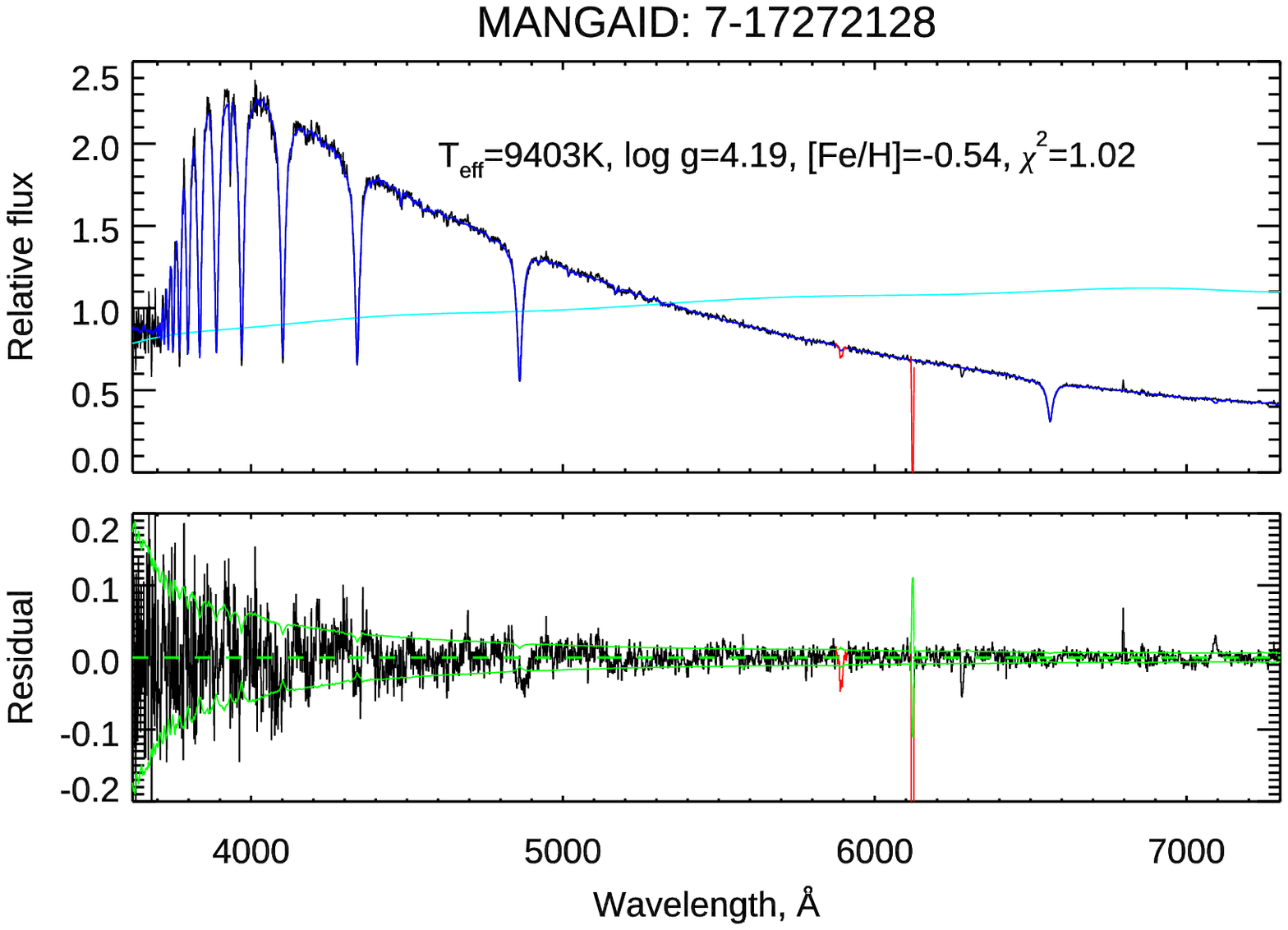}}
  \subfloat[ ]{ \includegraphics[scale=0.45,angle=0]{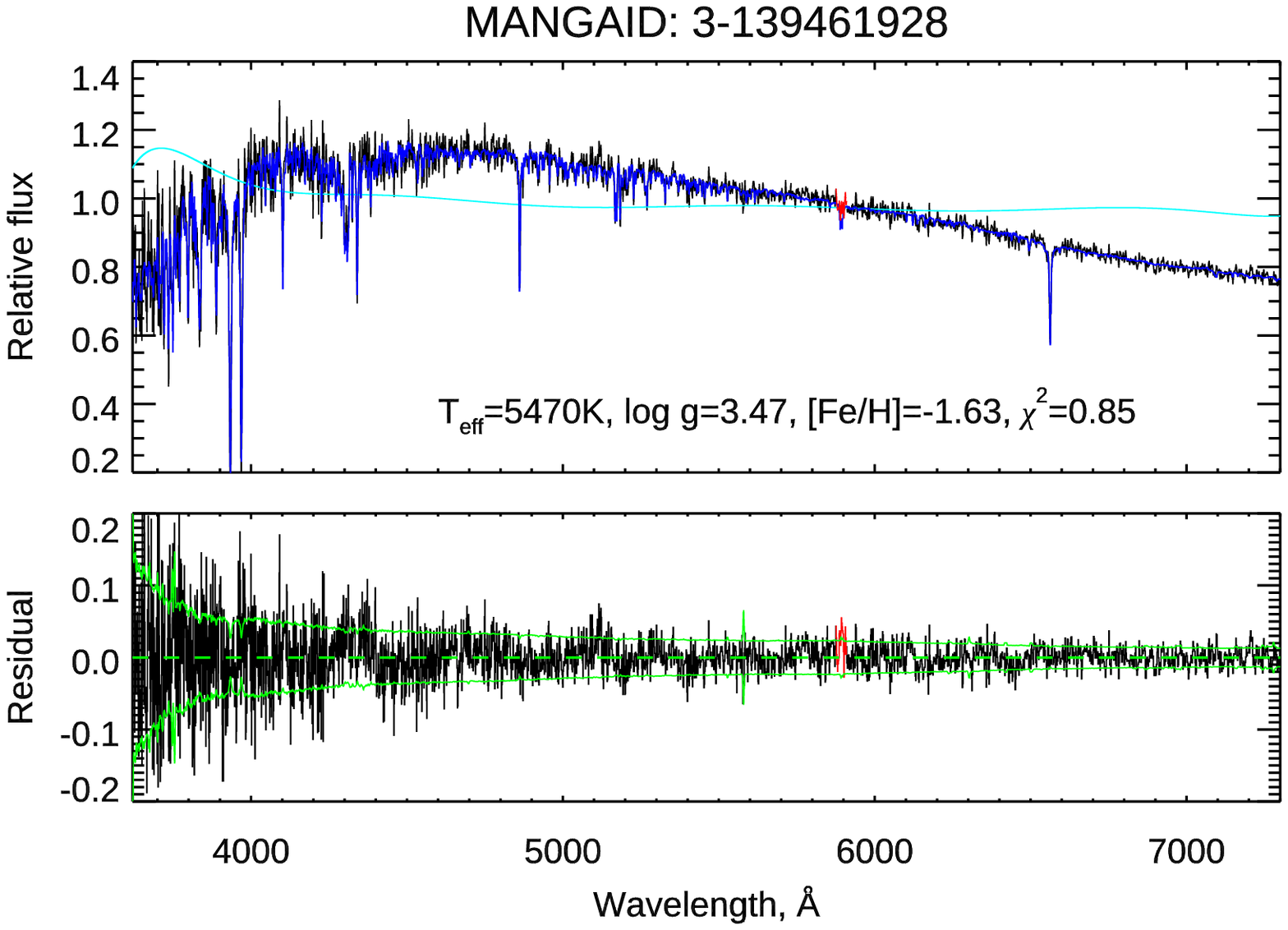}}  
\caption{Full-spectrum-fitting examples for MaStar with MANGA ID \text{27-1619},   \text{7-17427001},  \text{3-126081903},  \text{4-16745},  \text{7-17272128}, and  \text{3-139461928}, respectively. Upper panels: The data are shown as a black line,
the best-fit is shown as a blue line, 
and the polynomial is shown as a cyan line in each of the plots. Lower panels: Residuals between the data and the best-fit model, with the green line marking 
the one-sigma error from the data. Red pixels mark the bad pixels flagged in the data.  }
\label{fig:ulysample}
\end{figure*}

Figure~\ref{fig:ulysample} shows some examples of the full-spectrum fits to MaStar spectra. 
The black spectra are the observations from MaStar, the blue
spectra are the best fits from the MILES interpolator. The cyan lines show the polynomial used, and the red-line regions mask the bad
pixels. The bad pixels include pixels flagged by MaStar's mask flag and the region around the Na D 5900 feature, since the latter usually suffers
from interstellar absorption. 
The lower panel shows the residual between the data and the model, with the green spectrum marking the one-sigma error level. 
The derived stellar atmospheric parameters and reduced $\chi^2$ are provided in each of the panels. 

\section{Quality control }\label{secqc}

We fit 8646 
spectra measured for 3321 unique stars in MaStar. In order to make sure the stellar parameters derived from 
the full-spectrum-fitting process are sensible, 
we validated the spectral fitting on a case-by-case basis as needed.

\subsection{Data-Quality Effects}

We define two wavelength ranges to monitor the general signal-to-noise of the MaStar spectra -- blue: 3600--5500\AA\, 
and red: 5000--7500\AA. Wavelengths beyond 7500\AA\ are out of the range of the MILES templates, and therefore are not discussed here.   
Among the 8646 observations, around $0.4\%$   
have low signal-to-noise, less than five in the blue and red regions. If either of the above wavelength ranges has a median signal-to-noise 
value less than five, or the median signal-to-noise value over the entire wavelength range is less than 35,  
the spectra are tagged as having low signal-to-noise. 
We decided not to use their solutions due to these limitations.  

\subsection{Best-Fit Assessment} 

\begin{figure}
\centering
\includegraphics[scale=0.60,angle=0]
    {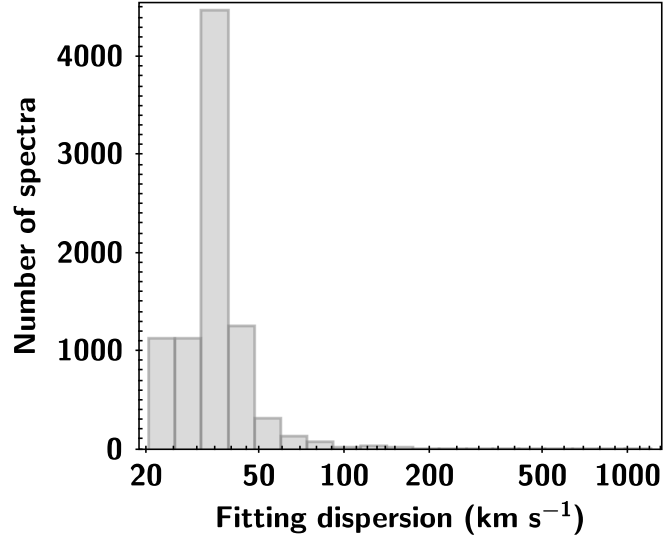}
\caption{Distribution of the fitted dispersion of the 8646 MaStar spectra in DR1. A total of 51
        out of 8646 spectra have fitted dispersions
         larger than 200  $\rm km s^{-1}$.}
\label{fig:fitdispdistri}
\end{figure}

\begin{figure}
\centering
\includegraphics[scale=0.60,angle=0]
    {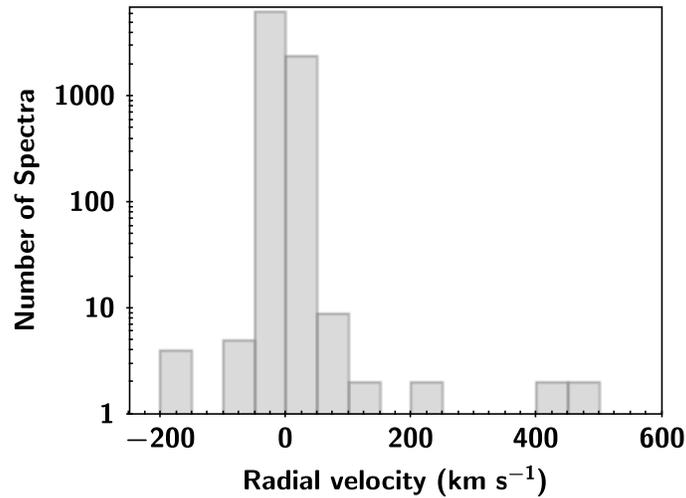}
\caption{Distribution of the fitted radial-velocity residuals of the MaStar spectra in DR1. A total of 6
         out of 8646 spectra have fitted 
         radial velocities larger than 400  $\rm km s^{-1}$.}
\label{fig:fitrvdistri}
\end{figure}

Since the interpolator itself has boundaries in the parameter space,
fits that return value at the extremes of the parameter space
at $\rm T_{eff}$=[2800, 40000]K, $\rm [Fe/H]=[-2.5,+1.0]$, $\log g= [0.0,5.9]$ (48 objects\footnote{These are dominated by giants and sub-giants whose metallicity could not be determined correctly.}) are rejected.
As a flexible fitting package, the fit process allows template broadening to mimic stellar rotation, however, we do not expect  
broadening that is comparable to galaxy velocity dispersions, i.e., a few hundred $\rm km\ s^{-1}$. Such large template broadening may 
indicate a template mismatch, low signal-to-noise ratio, or a possible binary. 
Either way, if the fitting results in a broadening larger than 200 $\rm km\ s^{-1}$, we classify it as unsuccessful.
The distribution of the fitting dispersion is shown in Figure~\ref{fig:fitdispdistri}.
The great majority of MaStar spectra are shifted to the rest-frame, therefore the radial-velocity fit we perform is to correct 
for any residual velocity offset. As a result, radial-velocity residuals
with large values (i.e., $rv \geq 400\ \rm km\ s^{-1}$) may also indicate 
unsuccessful fits. We show the distribution 
radial-velocity residuals in Figure~\ref{fig:fitrvdistri}.
Usually, when this happens the package confuses the absorption features between the data and the models. 
The corresponding stellar parameters are therefore not considered secure. 

\begin{figure}
\centering
\includegraphics[scale=0.60,angle=0]
    {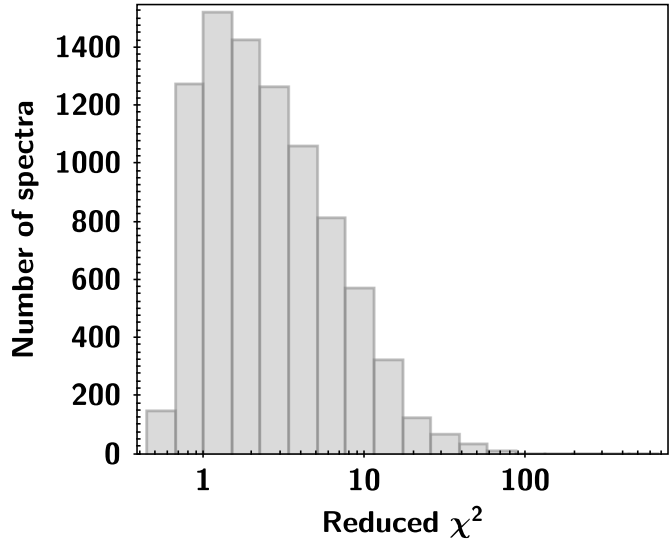}
\caption{Distribution of reduced $\chi ^2$ from the full-spectrum-fitting of the MaStar spectra in DR1. 
        10 out of 8646 fitting has reduced $\chi ^2$ larger than 100.}
\label{fig:fitchidistri}
\end{figure}

Apart from the above issues, we also had a few cases for which the derived errors of the corresponding parameters was zero.  
Those fits were rejected as failed fits as well.
The full-spectrum-fitting process calculates the reduced $\chi^2$, which is an
indicator of the success of the fit. We show the distribution of the reduced $\chi^2$ in Figure~\ref{fig:fitchidistri}.
We flagged the fits with reduced $\chi^2 \geq 150$. Normally this happens to input spectra with strong
emission features at very low ($\rm T_{eff} \lesssim 3700 K$) or very high ($\rm T_{eff} \gtrsim 30,000 K$)  temperatures.

Note that the original MILES library has a limited number of cool dwarfs \citep[$\sim$ 15 stars;][]{Yan19}. Thus, the results for cool stars  
based on this version of the interpolator need to be treated with caution.

In addition, many stars in MaStar DR1 have repeated observations. Independent measurement of their stellar parameters from full-spectrum-fitting gives consistent results, with the stellar parameters derived from our fitting of different spectra of the same object typically differing by
$\delta \rm T_{eff} \sim $40K, $\delta \log g \sim $ 0.04 dex, and $\delta \rm [Fe/H] \sim 0.05$ dex.

\section{Validation}\label{secval}

Because the MILES interpolator is constructed from 
the MILES library, it can only provide valid results in those regions of parameter space that are well-sampled  
by the MILES library. When it is used to fit stars that do not have a sufficient number of 
similar stars in the MILES library, it may produce results with very large errors. In fact, \citet[][see their Table\ 2]{Coelho20}  
reported $\sim 70$ stars in the MILES library which may not be suitable for stellar-population model inputs.
Here, we provide a way to identify such cases using only the resulting parameters and the training set for the interpolator. 
While this will not identify all large errors, it significantly reduces their rate. 
It also allows us to estimate the fraction of the remaining contamination of the derived stellar parameters due to the uneven input-parameter space in the interpolator.

This method assumes we can only reliably estimate the parameters for a particular stellar spectrum if its parameters are completely enclosed by the interpolator's training set.
However, we only know the output result from the interpolator. Thus, we establish a metric based on the output parameters of a star
and the input parameters of the training set. We then try to determine a threshold for this metric using the training set itself as a test sample. By fitting the training-set spectra, the MILES spectra, with the MILES interpolator itself, we can check if any spectra have large offsets in their derived parameters, and whether they can be successfully identified using the metric and the associated threshold.

First, we fit individual MILES spectra using the MILES interpolator. This is also a basic input-output test for the ULySS code and the MILES interpolator. 
In principle, the output stellar parameters should be identical or very close to the input values, but this is not always be case.

\begin{figure}
\centering
\includegraphics[scale=0.80,angle=0]
    {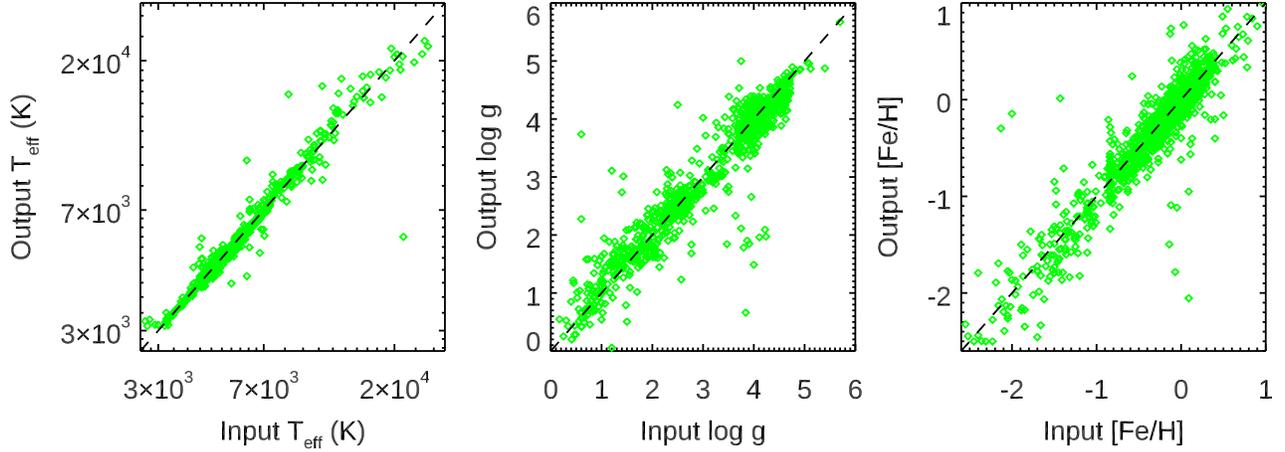}
\caption{Recovered stellar parameters for MILES spectra using the MILES interpolator versus their input parameters from \citet{Prugniel11}.
    The black dashed lines mark the one-to-one relations.
    }
\label{fig:milesone2one}
\end{figure}

\begin{figure*}[ht!]
\centering
\captionsetup[subfigure]{labelformat=empty}
  \subfloat[ ]{ \includegraphics[scale=0.50,angle=0]{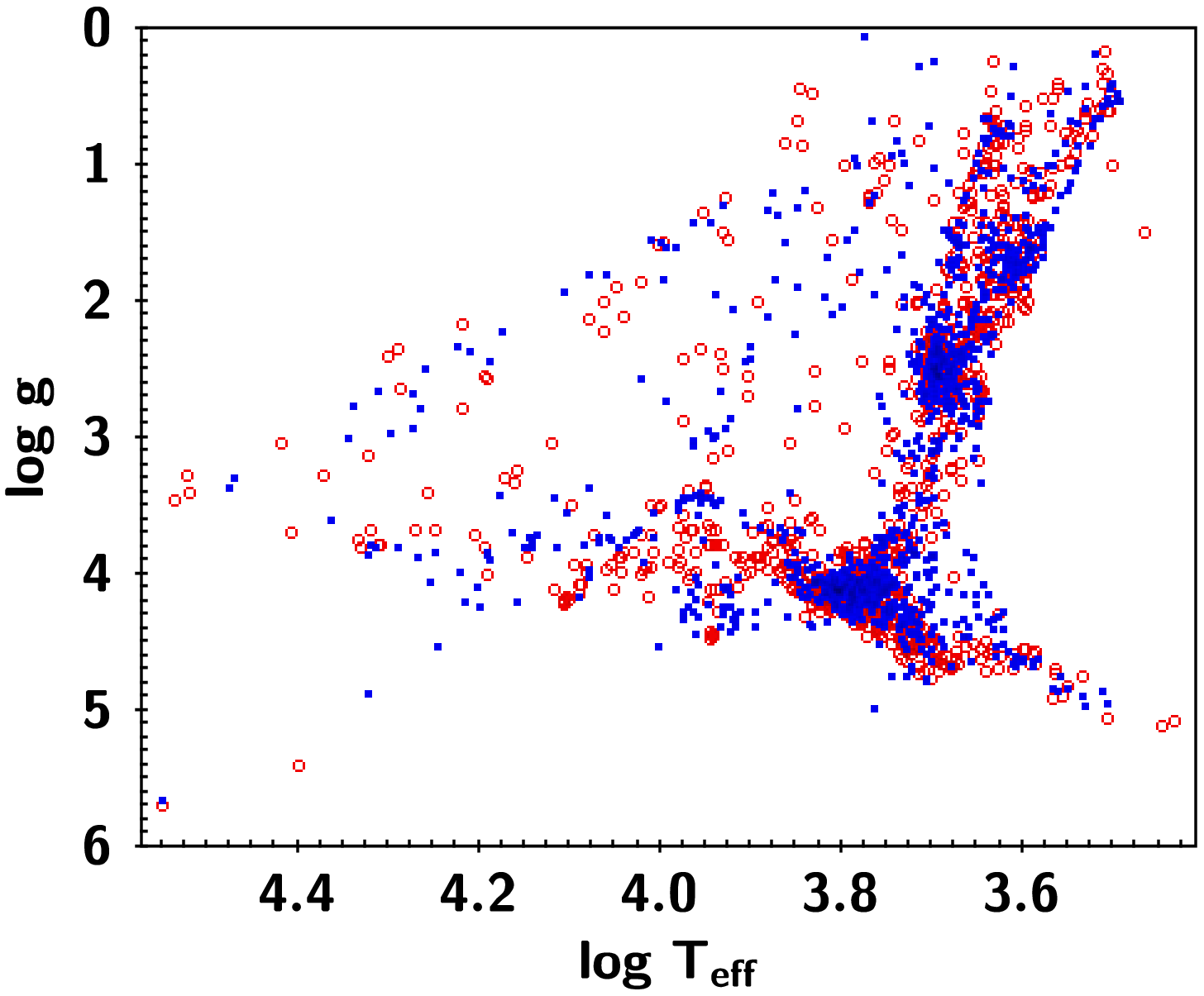}}
  \subfloat[ ]{ \includegraphics[scale=0.50,angle=0]{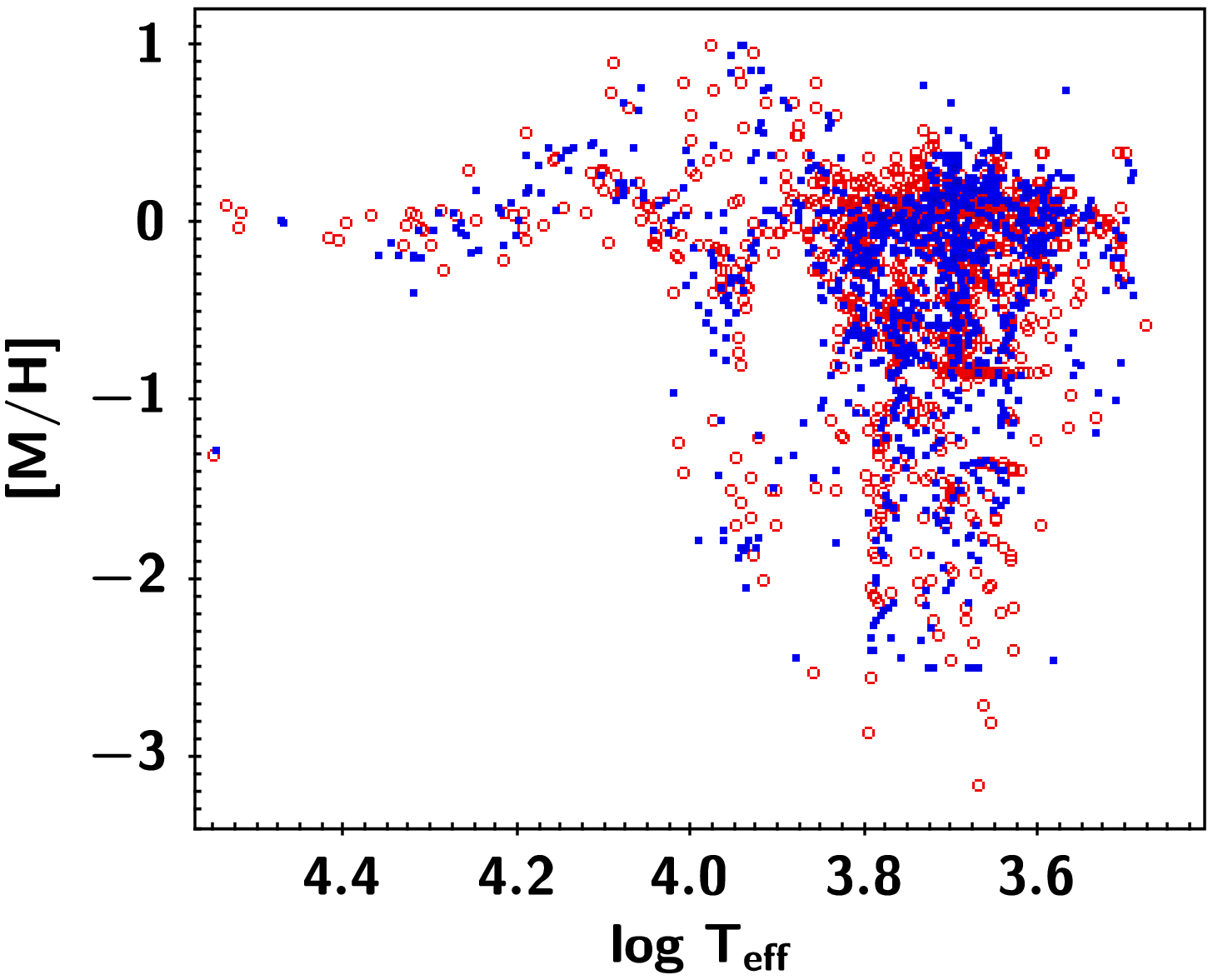}}
\caption{
Comparison between input (red open circles) and output (blue dots) stellar-parameter distributions of the MILES spectra. The `output' parameters are derived by fitting these MILES spectra with the MILES interpolator, which is built using the same set of spectra, and the `input' parameters based on \citet{Prugniel11}.  From inspection, 
the recovered parameters differ from the input set in several regions of the parameter space. 
}

\label{fig:milesrecover}
\end{figure*}

Figure~\ref{fig:milesone2one} shows the recovered 
stellar parameters for the MILES spectra compared with their input values.
Note that the calculated stellar parameters of MILES stars are not always the same as the input parameters as given by \citet{Prugniel11}. 
To inspect the global parameter distribution of the results, we show the parameter distribution of the input and output results in
Figure~\ref{fig:milesrecover}, where the red open circles are the stellar parameters from \citet{Prugniel11} and the blue solid dots are the 
recovered stellar parameters of the MILES spectra using the MILES interpolator. 
Due to the uneven distribution of the training set in the stellar-parameter space, 
the recovered parameters are biased towards regions 
with more stars in the input-parameter space. This is more apparent in the very hot or very cool temperature regions, where 
the interpolator's training set  has fewer stars.

\begin{figure}
\centering
\includegraphics[scale=0.60,angle=0]
    {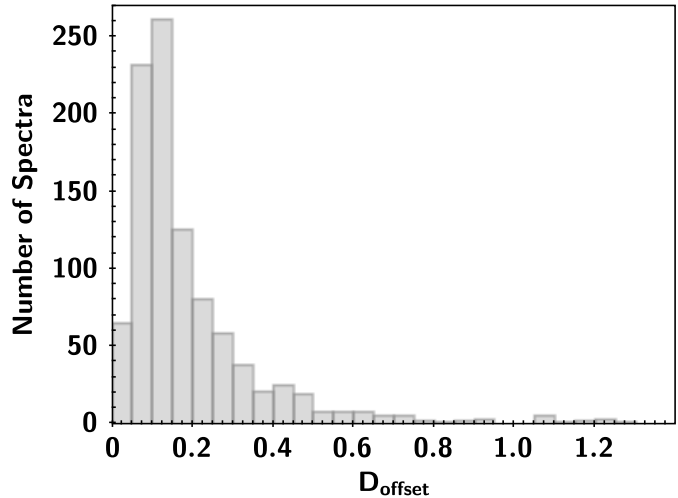}
\caption{Distribution of scaled distance $D_{\rm offset}$ (see text for details) of input parameters of the MILES interpolator based on \citet{Prugniel11}.}
\label{fig:milesparoff}
\end{figure}

\begin{figure*}
\centering
\captionsetup[subfigure]{labelformat=empty}
\subfloat[ ]{  \includegraphics[scale=0.50,angle=0]{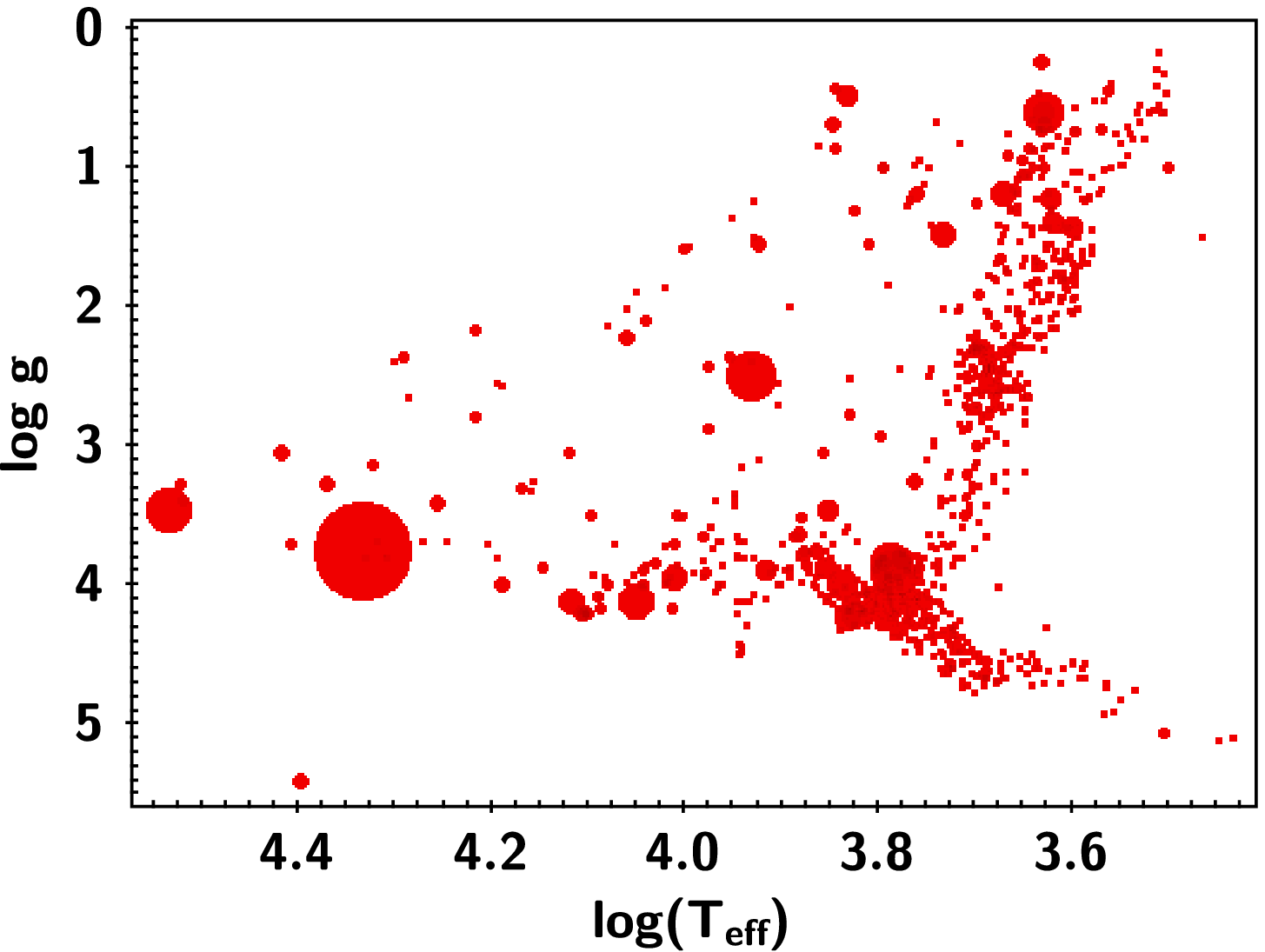}}
\subfloat[ ]{ \includegraphics[scale=0.50,angle=0]{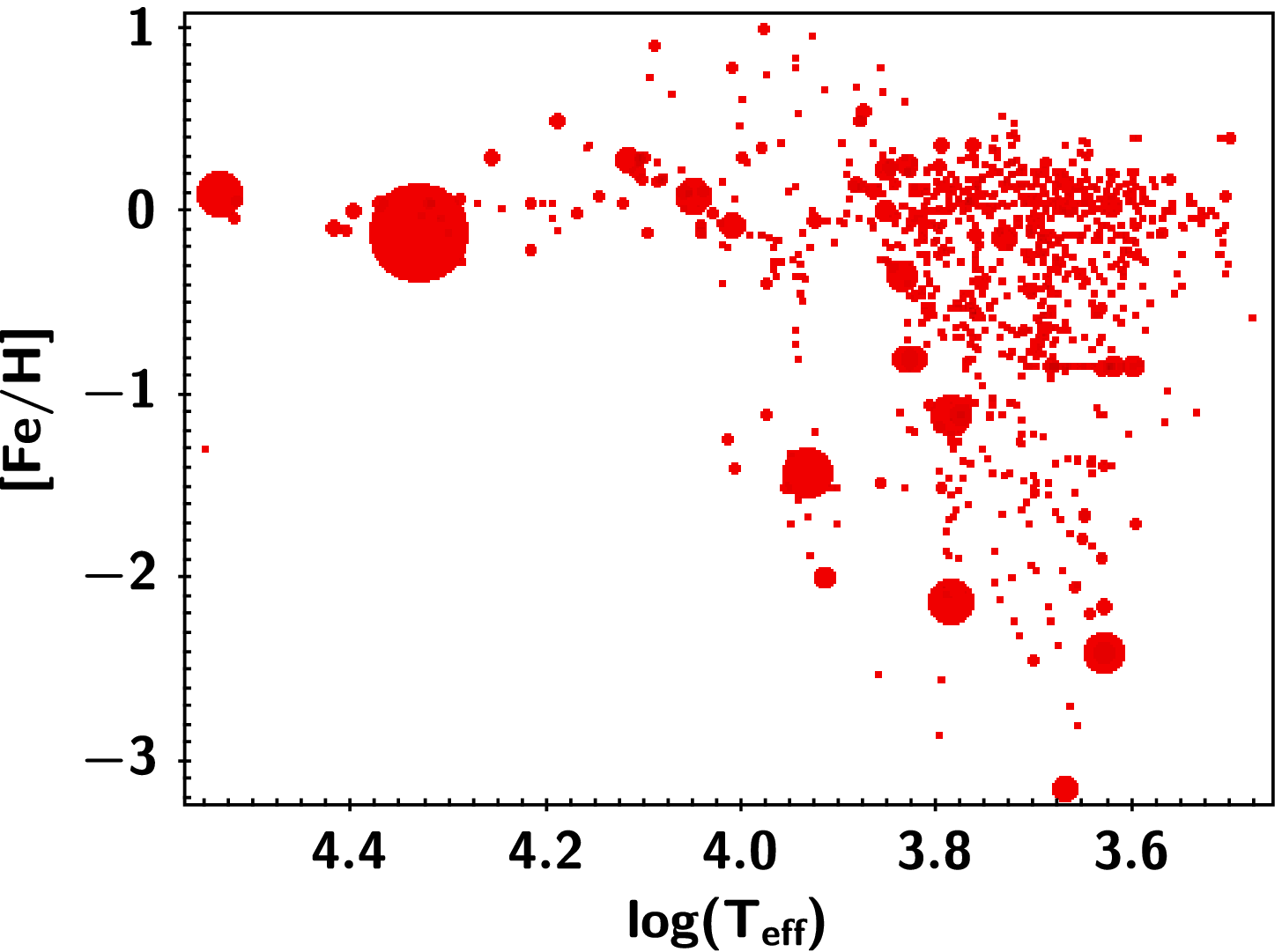}}
\caption{Stellar-parameter distributions of the input MILES spectra. The symbol size is scaled to  $D_{\rm offset}$; larger 
symbols correspond to larger systematic errors in recovering the stellar parameters.}
\label{fig:milesrecoveroffs}
\end{figure*}

The three dimensions of the parameter space ($\log T_{\rm eff}$,  $\log g $, and ${\rm [Fe/H]}$) have different units. We define a normalized three-dimensional distance unit between any two points in the parameter space  by normalizing the difference in each dimension appropriately so that the typical uncertainty in each dimension is about equal in this normalized unit. This can be approximately achieved if we scale the maximum range spanned in $\log T_{\rm eff}$ (from 3.477 to 4.550) to 10, scale that in $\log g$ (from 0.17 to 5.7) to 3, and scale that in [Fe/H] (from -3.15 to +1.0) to 2. For each MILES star, we compute the distance between its recovered parameters and its input parameters in this normalized unit, and call it $D_{\rm offset}$. We show the distribution of $D_{\rm offset}$ in Figure~\ref{fig:milesparoff}. A total of 94\% of the stars have $D_{\rm offset}$ less than 0.5. 
Figure~\ref{fig:milesrecoveroffs} illustrates the stellar-parameter stability. Regions with larger symbols are the ones with larger uncertainties in the input-parameter space, i.e., larger $D_{\rm offset}$. 
If a $D_{\rm offset}$ is shared evenly among the three dimensions, the offset in each dimension is 0.29 in the normalized unit, which corresponds to $\Delta\log T_{\rm eff}=0.03$, $\Delta \log g=0.5$, and $\Delta \rm [Fe/H] = 0.6$.

For each star, given its {\it recovered} parameters, we look for its neighbors in the (input) parameter space among the training set. Note that the neighbor search is between the {\it output} parameters of the star of interest and the {\it input} parameters of the training set. We do it this way because we would only know the output parameters for MaStar objects.  For each star (MILES or MaStar), we define the 4th-nearest-neighbor distance among the training set, again in the normalized unit, and call it $D_{\rm x4NN}$. We choose to use the four nearest stars because 
a minimum of four points are needed to define a volume (a tetrahedron) that encloses a point in the 3D parameter space.
We aim to find a threshold in $D_{\rm x4NN}$ to ensure that most stars with $D_{\rm x4NN}$ smaller than the threshold have small $D_{\rm offset}$. 

\begin{figure}
\centering
\includegraphics[scale=0.50,angle=0]
    {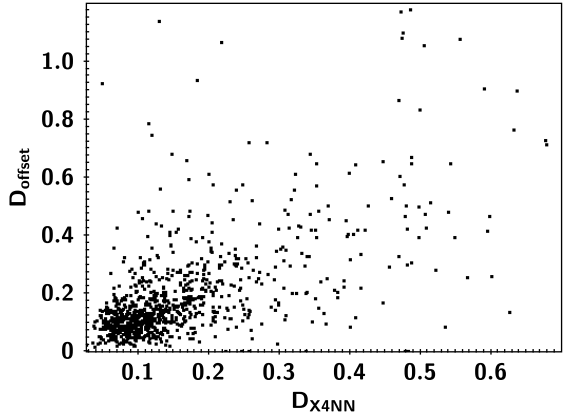}
\caption{$D_{\rm offset}$ as a function of $D_{\rm x4NN}$. Note that distances are scaled (see text for details).} 
\label{fig: Doffset_Dx4NN}
\end{figure}

There could be cases in which the four nearest neighbors around the star of interest are all clustered on one side, and are much closer to each other than to the star of interest. Here we do not require the star of interest to be completely enclosed by the tetrahedron, as some moderate extrapolation could still give reasonable results. We only require that the $D_{\rm x4NN}$ be smaller than the largest pair-wise distance among its neighbors. 
There are six distances among the four neighbors. We call the maximum among them $D_6$. We require $D_{\rm x4NN}$ to be smaller than $D_6$. 

Figure~\ref{fig: Doffset_Dx4NN} shows $D_{\rm offset}$ vs. $D_{\rm x4NN}$ for all MILES stars. We highlight in this figure those stars with $D_{\rm x4NN}$ larger than their respective $D_6$, which are excluded from the following statistical calculations. 

\begin{figure}
\centering
\includegraphics[scale=0.50,angle=0]
    {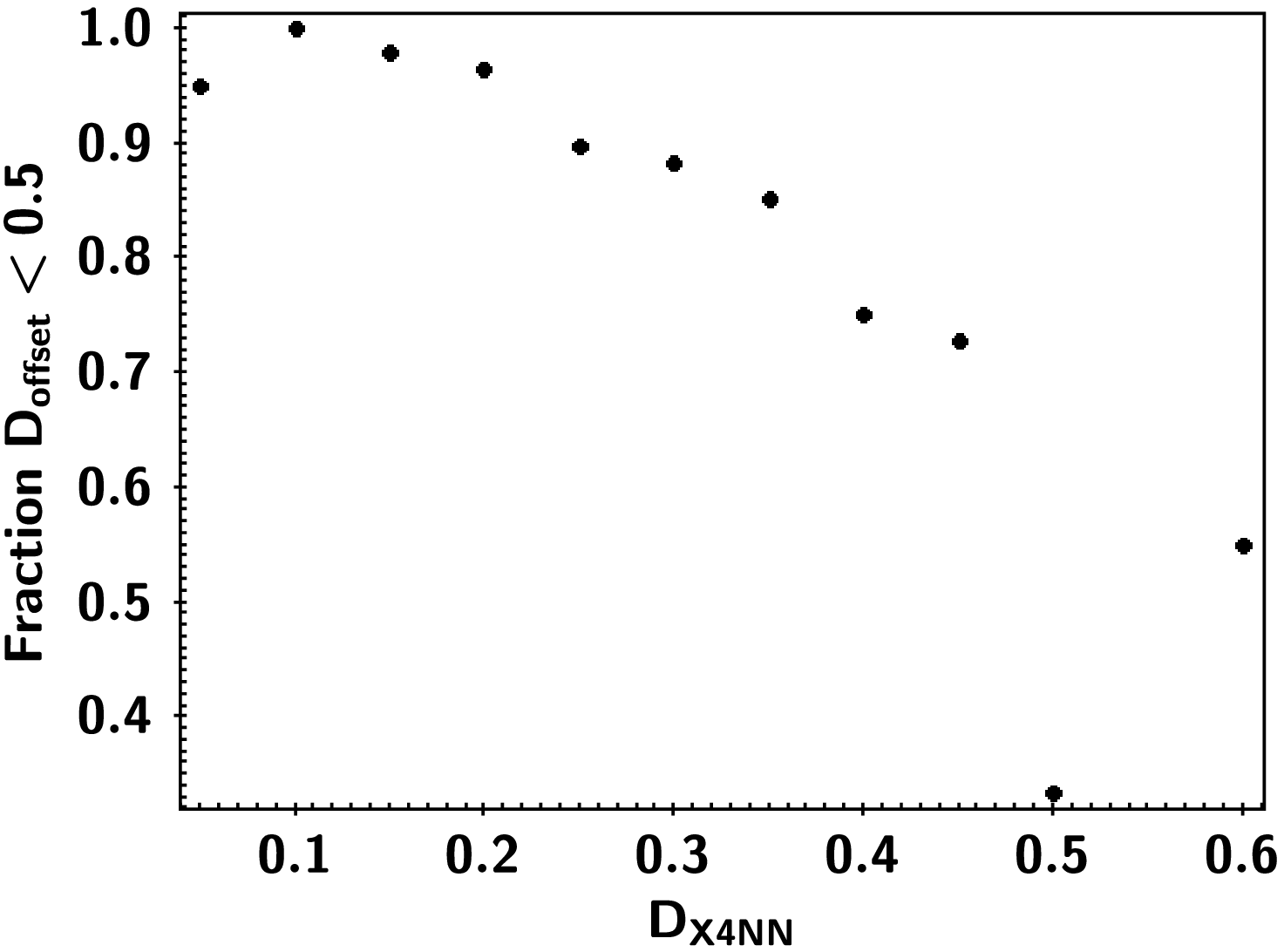}
\includegraphics[scale=0.50,angle=0]
    {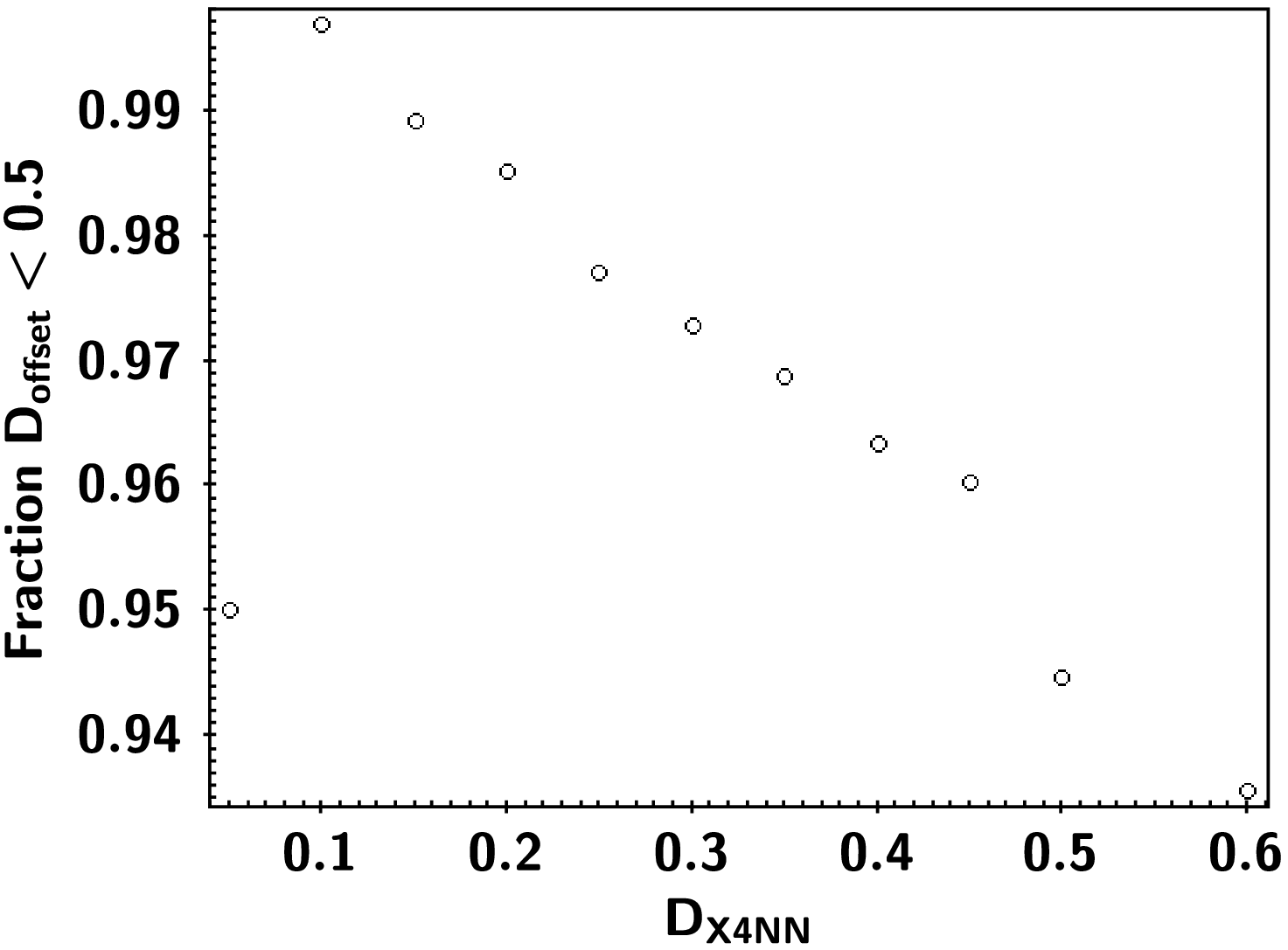}  
\caption{Left: Fraction of MILES stars in bins of $D_{\rm x4NN}$ that have $D_{\rm offset}<0.5$.  Right: Fraction of MILES stars with $D_{\rm x4NN}$ less than a threshold that have $D_{\rm offset}<0.5$. Stars with $D_{\rm x4NN}>D_6$ are excluded from the statistical calculations.}
\label{fig:fraction_vs_threshold}
\end{figure}

We consider the parameter measurements for a star to be reasonable when its $D_{\rm offset}$ is less than 0.5. In order to determine the threshold for the $D_{\rm x4NN}$ metric, we compute the fraction of MILES stars with $D_{\rm offset}<0.5$  among all those with $D_{\rm x4NN}$ less than a threshold, while excluding those with $D_{\rm x4NN}>D_6$. 
Figure~\ref{fig:fraction_vs_threshold} shows
this fraction as a function of varying threshold. From this plot, we find that when we adopt a threshold of $D_{\rm x4NN}< 0.35$, 96.9\% of the stars have $D_{\rm offset}<0.5$.

\begin{figure*}[ht!]
\centering
\captionsetup[subfigure]{labelformat=empty}
\subfloat[ ]{ \includegraphics[scale=0.55,angle=0]{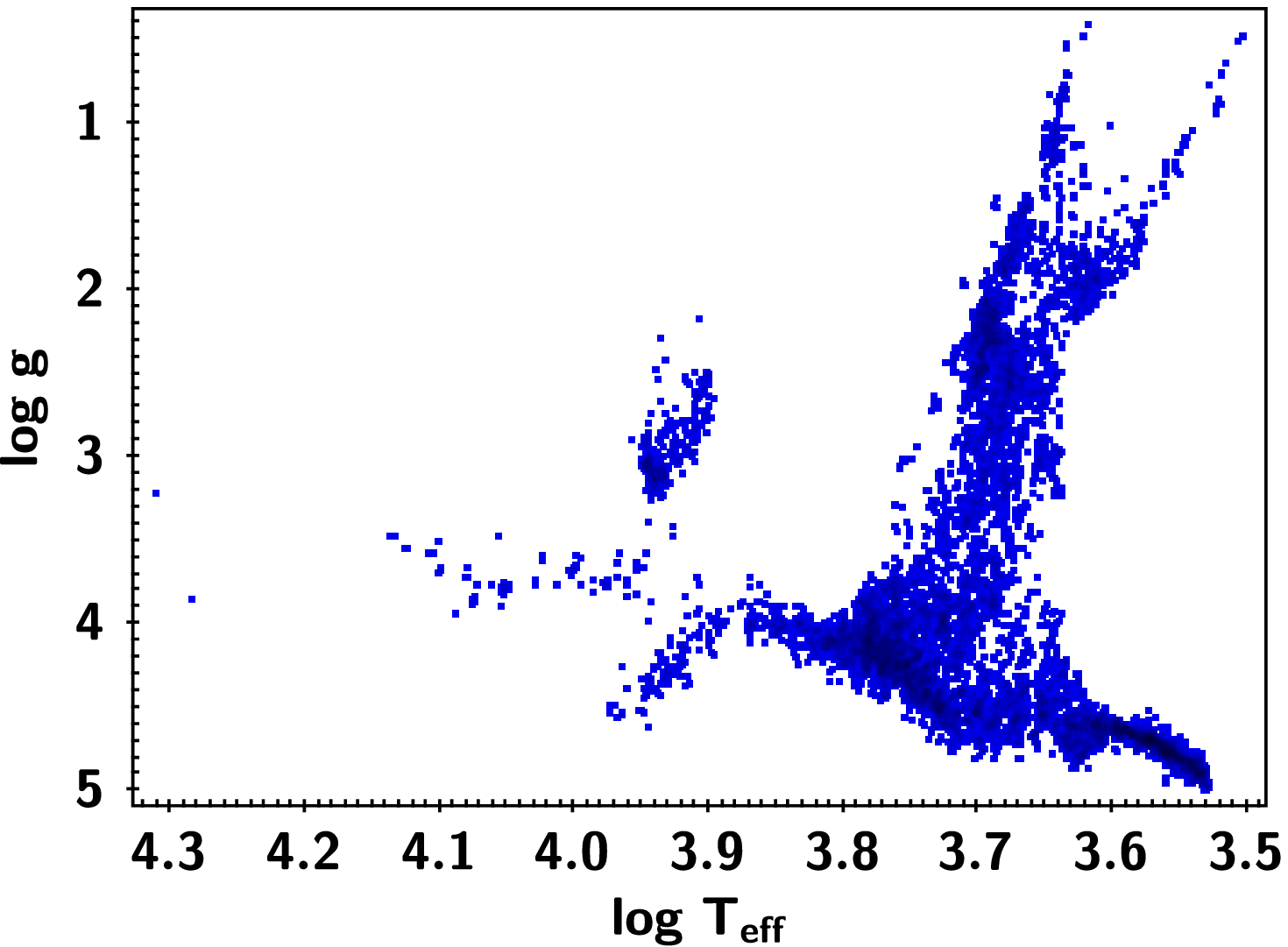}}
\subfloat[ ]{ \includegraphics[scale=0.55,angle=0]{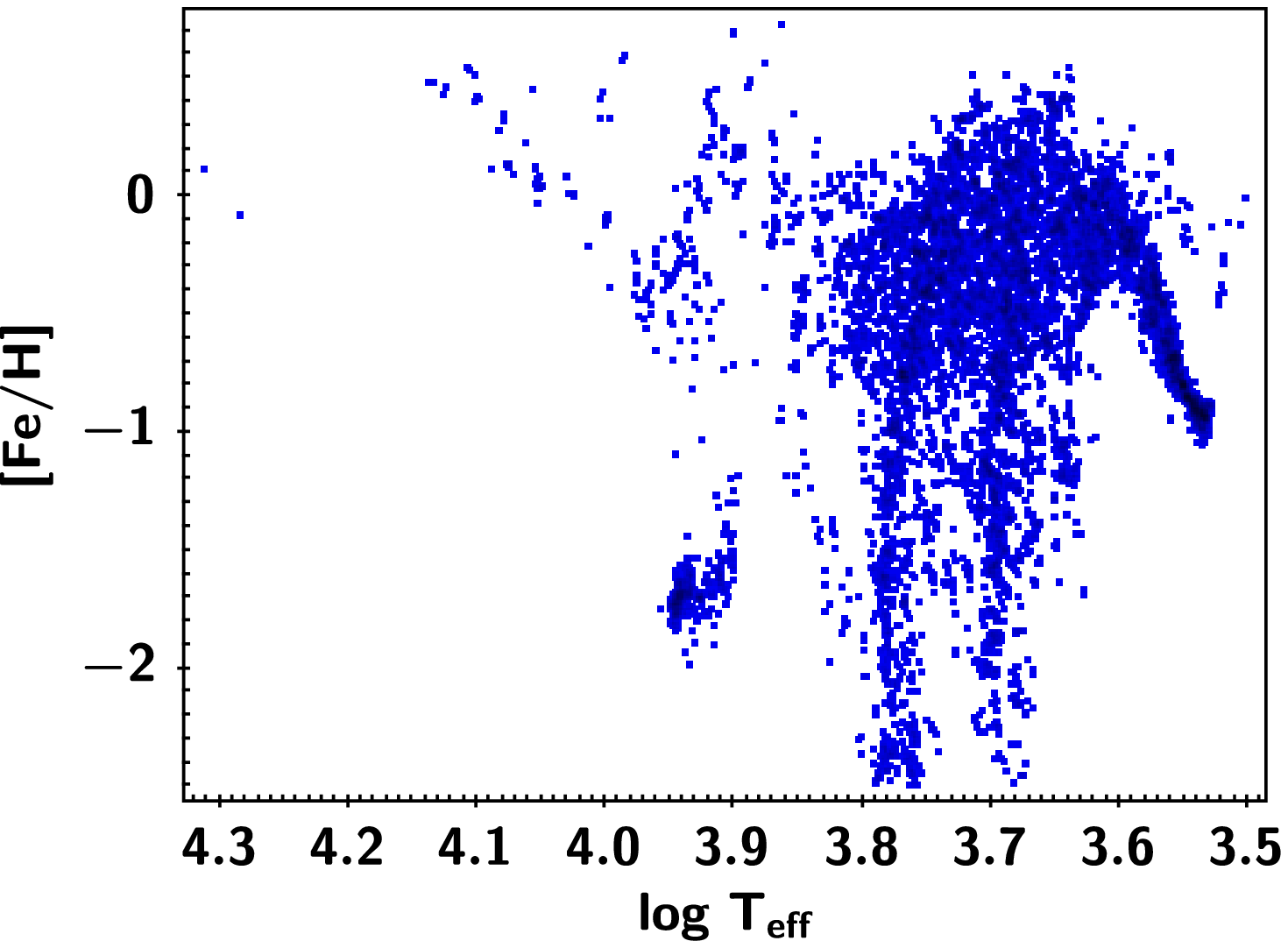}}
\caption{Adopted stellar-parameter distribution of MaStar from this work.
}
\label{fig:mastarparams}
\end{figure*}

With the threshold defined by $D_{\rm x4NN}$, we can assess the reliability of the parameter measurement for MaStar. Among the 8646 MaStar spectra, 87\% have $D_{\rm x4NN} <0.35$, $D_{\rm x4NN}< D_6$, and pass the quality control test. We consider these cases to have reliable stellar-parameter estimates. 
Given the test based on MILES, we expect 96.9\% of these cases would have a parameter offset less than 0.5 in the normalized unit. We show their stellar-parameter distribution in Figure~\ref{fig:mastarparams}. 
Table~\ref{sampletable} lists these values, so that they can be compared with either literature values or other methods.

\section{Consistency with input parameters}\label{secexqc}

Apart from the above internal quality control, we can also validate the parameters we derived from this work by checking their consistency 
with the input catalog stellar parameters.
As mentioned in \citet{Yan19}, we selected stars from existing stellar-parameter catalogs, including ASPCAP uncalibrated atmospheric parameters \citep{Garcia16, Holtzman18, Jonsson18} from the latest SDSS-IV Data Release \citep[APOGEE DR16;][]{Ahumada19}, Sloan Extension for Galactic Understanding and Exploration \citep[SEGUE;][]{yanny09} , and the Large Sky Area Multi-Object Fiber Spectroscopic Telescope \citep[LAMOST;][]{cui12, deng12, zhao12}. Because the stellar parameters from these above projects were derived using different methods, 
it is important to address the parameter consistency.

\begin{figure}
\centering
\includegraphics[scale=0.80,angle=0]
    {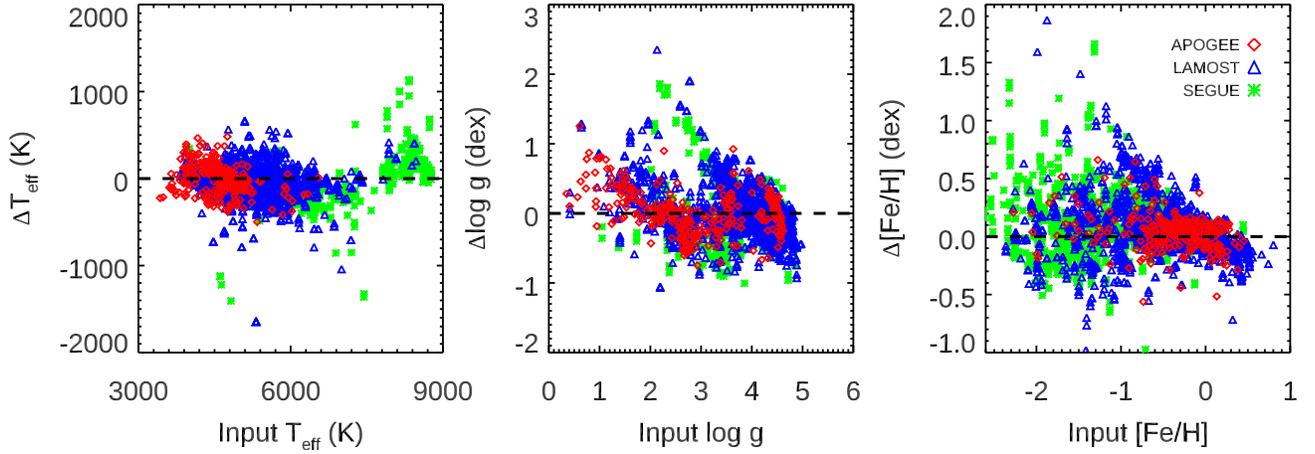}
\caption{Stellar parameter derived from this work compared with the input stellar parameters from APOGEE, SEGUE, and LAMOST.
In each panel we show the differences between the stellar parameters from this work and the input parameters. }
\label{fig:paramcmp}
\end{figure}

We have identified 2460 spectra that passed the quality thresholds and have APOGEE parameters, where 288 of them are in the fine parameter category in APOGEE \citep{Garcia16, Holtzman18, Jonsson18, Ahumada19}, 
1064 such spectra which have SEGUE parameters, and 1810 spectra which have LAMOST parameters. 
We show the stellar parameters from this work compared with these literature parameters in Figure~\ref{fig:paramcmp}. 
The mean $\rm T_{eff}$ determined in this work are about 50 K cooler than these literature values, with an rms of 163 K; 
the mean  $\log g$ determined in this work are about 0.03 dex lower than the input literature values, with an rms of 0.30 dex; 
the mean  $\rm [Fe/H]$ determined in this work are about 0.06 dex lower than the input literature values, with an rms of 0.20 dex.
These values are typically consistent with the accuracy of ULySS's parameters \citep[e.g.,][]{Prugniel11}, except for a larger 
deviation in $\rm T_{eff}$, which is likely due to the multiple input methods from different sources. 
Besides, these accuracies are within the range required by robust population synthesis \citep[100K,][]{Maraston05}.

We also explore the stellar-parameter consistency from different input-parameter sources. 
A mean negative offset of 9 K is found in comparing with $\rm T_{eff}$ from APOGEE, with an rms of 174 K;
a mean negative offset of 71 K is found in comparing with $\rm T_{eff}$ from SEGUE, with an rms of 167 K;
a mean negative offset of 41 K is found in comparing with $\rm T_{eff}$ from LAMOST, with an rms of 154 K.
A mean offset of 0.01 dex is found in comparing with $\log g$ from APOGEE, with an rms of 0.32 dex;
a mean negative offset of 0.07 dex is found in comparing with $\log g$ from SEGUE, with an rms of 0.29 dex;
a mean negative offset of 0.01 dex is found in comparing with  $\log g$ from LAMOST, with an rms of 0.29 dex.
A mean negative offset of 0.03 dex is found in comparing with $\rm [Fe/H]$ from APOGEE, with an rms of 0.13 dex;
a mean positive offset of 0.10 dex is found in comparing with $\rm [Fe/H]$ from SEGUE, with an rms of 0.23 dex;
a mean positive offset of 0.04 dex is found in comparing with $\rm [Fe/H]$ from LAMOST, with an rms of 0.19 dex.
In general, given the rms as typical accuracy of ULySS's parameters, our results agree well with the input stellar parameters. 

We acknowledge that due to the available template interpolator \citep[i.e.,][]{Prugniel11}, which does not contain the $[\alpha/Fe]$ dimension, we are not providing $[\alpha/Fe]$ in this work.  
Our team is working on different methods of estimating the stellar parameters. The $[\alpha/Fe]$ dimension will be addressed in future work.

\section{Conclusions}\label{secconclusion}

We have examined 8646 spectra from the first release of MaStar to estimate their basic stellar parameters, using the MILES interpolator
\citep{Prugniel11} and the full-spectrum-fitting package ULySS. 
We carried out a sanity check of this fitting method by feeding the same spectra used to build the interpolator through the fitting algorithm. 
This input-output test shows that the algorithm is not always robust. We defined a set of criteria to identify regions of parameter space where it is safe to apply this fitting method, and evaluated the purity of the resulting sample. We present the parameters for a subset of MaStar spectra that reside within the safe region of the parameter space identified. 
Our `secure' stellar parameters cover a wide range of the parameter space:  $\rm 3179 \leq T_{eff} \leq 20517 K$,
surface gravity $0.40 \leq \log g \leq 5.0$, and metallicity $-2.49 \leq \rm[Fe/H] \leq +0.73$.
For stars in common with the APOGEE, SEGUE, and LAMOST surveys, we compared the results with the derived parameters by these surveys and found good agreement: the mean differences of $\rm T_{eff}$ determined in this work are about 50 K cooler than the 
input $\rm T_{eff}$, with an rms of 163 K;  the mean differences of  $\log g$ determined in this work are about 0.03 dex lower than the input 
$\log g$, with an rms of 0.30 dex; the mean differences of $\rm [Fe/H]$ determined in this work are about 0.06 dex lower than the input 
values, with an rms of 0.20 dex.
Note that this work is based on the interpolator of MILES stellar library. MaStar objects that are outside of the MILES parameter
space may not have secure estimates of their stellar parameters. Therefore, we only list the parameters that passed the quality control tests.
These parameters have been already used for the calculation of the first MaStar-based stellar-population models
\citep{Maraston20}. The models have been tested with Milky Way globular cluster spectra, obtaining good results, providing reassurance
of the overall quality of these parameters \citep[see details in][]{Maraston20}.

\begin{table}
\centering
\caption{Stellar Parameters Adopted in this Work}
 \begin{threeparttable}\label{sampletable}
 \footnotesize 
\begin{tabular}{|l|r|r|r|r|r|r|r|r|r|}
\hline
  \multicolumn{1}{|c|}{MANGAID} &
  \multicolumn{1}{c|}{RA}  &
  \multicolumn{1}{c|}{DEC}  &
  \multicolumn{1}{c|} {$\rm T_{eff}$} &
  \multicolumn{1}{c|}{$\log g$ } &
  \multicolumn{1}{c|}{$\rm [Fe/H]$} &
  \multicolumn{1}{c|}{$\rm T_{eff}$ err} &
  \multicolumn{1}{c|}{$\log g$ err} &
  \multicolumn{1}{c|}{$\rm [Fe/H]$ err} &
  \multicolumn{1}{c|}{$\chi^2$} \\
 & (deg) &(deg) &(K) &(cgs) &  \  &(K) &(cgs) & (dex) &  \\
\hline
  7-17152035 & 288.1182 & 51.4476 & 5124 & 2.31 & $-$1.36 & 5 & 0.01 & 0.01 & 5.07\\
  7-17443125 & 290.2698 & 51.3453 & 4502 & 2.15 & $-$0.38 & 4 & 0.01 & 0.01 & 3.34\\
  7-17373411 & 290.1849 & 49.8035 & 4839 & 2.33 & $-$0.29 & 3 & 0.01 & 0.00 & 6.16\\
  7-17182569 & 288.5502 & 49.5116 & 4855 & 3.18 & 0.30 & 4 & 0.01 & 0.00 & 4.00\\
  7-17182569 & 288.5502 & 49.5116 & 4868 & 3.26 & 0.29 & 6 & 0.02 & 0.01 & 2.28\\
  7-17113356 & 287.2459 & 51.0213 & 4797 & 2.21 & $-$0.65 & 3 & 0.01 & 0.00 & 5.03\\
  7-17372978 & 289.7807 & 50.2919 & 4583 & 1.59 & $-$0.99 & 6 & 0.02 & 0.01 & 2.94\\
  7-16750617 & 282.3300 & 43.3356 & 4144 & 1.90 & $-$0.23 & 1 & 0.01 & 0.00 & 9.48\\
  7-16565620 & 282.0194 & 44.5682 & 5451 & 3.43 & 0.12 & 4 & 0.01 & 0.01 & 4.48\\
  7-16557545 & 281.6392 & 44.0813 & 3840 & 1.61 & $-$0.03 & 1 & 0.00 & 0.00 & 14.38\\
\hline\end{tabular}
   \begin{tablenotes}
      \small
      \item A sample of the stellar parameters from this work. The full table is available in digital format only. 
    \end{tablenotes}
  \end{threeparttable}

\end{table}

\acknowledgments
\section*{acknowledgments}

This project made use of data taken in SDSS-IV. 

Funding for the Sloan Digital Sky Survey IV has been provided by the Alfred P. Sloan Foundation, the U.S. Department of Energy Office of Science, and the Participating Institutions. SDSS-IV acknowledges
support and resources from the Center for High-Performance Computing at the University of Utah. The SDSS web site is www.sdss.org.

SDSS-IV is managed by the Astrophysical Research Consortium for the 
Participating Institutions of the SDSS Collaboration including the 
Brazilian Participation Group, the Carnegie Institution for Science, 
Carnegie Mellon University, the Chilean Participation Group, the French Participation Group, Harvard-Smithsonian Center for Astrophysics, 
Instituto de Astrof\'isica de Canarias, The Johns Hopkins University, Kavli Institute for the Physics and Mathematics of the Universe (IPMU) / 
University of Tokyo, the Korean Participation Group, Lawrence Berkeley National Laboratory, 
Leibniz Institut f\"ur Astrophysik Potsdam (AIP),  
Max-Planck-Institut f\"ur Astronomie (MPIA Heidelberg), 
Max-Planck-Institut f\"ur Astrophysik (MPA Garching), 
Max-Planck-Institut f\"ur Extraterrestrische Physik (MPE), 
National Astronomical Observatories of China, New Mexico State University, 
New York University, University of Notre Dame, 
Observat\'ario Nacional / MCTI, The Ohio State University, 
Pennsylvania State University, Shanghai Astronomical Observatory, 
United Kingdom Participation Group,
Universidad Nacional Aut\'onoma de M\'exico, University of Arizona, 
University of Colorado Boulder, University of Oxford, University of Portsmouth, 
University of Utah, University of Virginia, University of Washington, University of Wisconsin, 
Vanderbilt University, and Yale University.

T. C. B. acknowledges partial support from grant PHY 14-30152 (Physics Frontier Center/JINA-CEE), awarded by the U.S. National Science Foundation.

{J.G.F-T} is supported by FONDECYT No. 3180210 and Becas Iberoam{\'e}rica Investigador 2019, Banco Santander Chile.

The research of J. D G. is also supported by NYU Abu Dhabi Grant AD022.
Y. C. acknowledges the support of NYU Abu Dhabi AD013 and NYU Abu Dhabi Grant AD022.


\begin{thebibliography}{}

\bibitem[Aguado et al.(2019)]{Aguado19}{Aguado}, D.~S., {Ahumada}, R., {Almeida}, A., et al. 2019, ApJS, 240, 23A
\bibitem[Ahumada et al.(2019)]{Ahumada19}{{Ahumada}, Romina,  {Allende Prieto}, Carlos, {Almeida}, Andres et al., 2019, arXiv, 191202905A}
\bibitem[Allen(1973)]{Allen73}{Allen}, C.~W., 1973, asqu.book, A
\bibitem[Beifiori et al.(2011)]{Beifiori11}{Beifiori}, A., {Maraston}, C., {Thomas}, D., {Johansson}, J., 2011, A\&A, 531A, 109B
\bibitem[Blanton et al.(2017)]{Blanton17} Blanton, M. R., Bershady, M. A., Abolfathi, B., et al. 2017, AJ, 154, 28B
\bibitem[Bohlin et al.(2017)]{BOSZ}Bohlin, R. C., {M{\'e}sz{\'a}ros}, S., Fleming, S. ~W., et al. 2017, AJ, 153, 234B
\bibitem[Bundy et al.(2015)]{Bundy15}{Bundy}, Kevin, {Bershady}, Matthew A., {Law}, David R., et al., 2015, ApJ, 798, 7B
\bibitem[Cardelli et al.(1989)]{Cardelli89}{Cardelli}, J.~A., {Clayton}, G.~C., {Mathis}, J.~S., 1989, ApJ, 345, 245C
\bibitem[Casagrande et al.(2008)]{Casagrande08}{Casagrande}, L., {Flynn}, C., {Bessell}, M., 2008, MNRAS, 389, 585C
\bibitem[Cenarro et al.(2001)]{Cenarro01}Cenarro, A. J., Cardiel, N., Gorgas, J., et al., 2001, MNRAS, 326, 959,
\bibitem[Cenarro et al.(2007)]{Cenarro07}{Cenarro}, A.~J., {Peletier}, R.~F., {S{\'a}nchez-Bl{\'a}zquez}, P., et al. 2007, MNRAS, 374, 664C
\bibitem[Chambers et al.(2016)]{pansref}Chambers, K. C., Magnier, E. A., Metcalfe, N., et al. 2016, ArXiv: 1612.05560
\bibitem[Chen et al.(2014)]{XSL}Chen, Yan-Ping, Trager, S. C., Peletier, R. F., et al. 2014, A\&A, 565A,117C
\bibitem[Coelho et al.(2005)]{Coelho05}{Coelho}, P., {Barbuy}, B., {Mel{\'e}ndez}, J., {Schiavon}, R.~P., {Castilho}, B.~V., 2005, A\&A, 443, 735C
\bibitem[Coelho et al.(2007)]{Coelho07}Coelho, P., Bruzual, G., Charlot, S., et al., 2007, MNRAS, 382, 498
\bibitem[Coelho et al.(2020)]{Coelho20}Coelho, P. R. T., Bruzual, G., \& Charlot, S., 2020, MNRAS, 491, 2025C
\bibitem[Conroy \& {van Dokkum}(2012)]{Conroy12}{Conroy}, Charlie, {van Dokkum}, Pieter, 2012, ApJ, 747, 69C
\bibitem[Cui et al.(2012)]{cui12}Cui, X.-Q., Zhao, Y.-H., Chu, Y.-Q., et al. 2012, Research in Astronomy and Astrophysics, 12, 1197, doi: 10.1088/1674-4527/12/9/003
\bibitem[de Laverny et al.(2012)]{deLaverny12}de Laverny, P., Recio-Blanco, A., Worley, C. C.,, Plez, B. 2012, A\&A, 544, A126
\bibitem[Deng et al.(2012)]{deng12}Deng, L.-C., Newberg, H. J., Liu, C., et al. 2012, Research in Astronomy and Astrophysics, 12, 735, doi: 10.1088/1674-4527/12/7/003
\bibitem[Diaz et al.(1989)]{Diaz89}{Diaz}, A.~I.,  {Terlevich}, E., {Terlevich}, R., 1989, MNRAS, 239, 325D
\bibitem[Falc{\'o}n-Barroso et al.(2011)]{Falcon-barroso11}{Falc{\'o}n-Barroso}, J., {S{\'a}nchez-Bl{\'a}zquez}, P., {Vazdekis}, A., et al., 2011, A\&A, 532A, 95F
\bibitem[Garc{\'{\i}}a P{\'e}rez et al.(2016)]{Garcia16}{Garc{\'{\i}}a P{\'e}rez}, A.~E., {Allende Prieto}, C., {Holtzman}, J.~A.,
  {et~al.} 2016, AJ 151, 144G
\bibitem[Gregg et al.(2006)]{ngsl}Gregg, M. D., Silva, D., Rayner, J., et al. 2006, hstc, conf, 209G
\bibitem[Gustafsson et al.(2008)]{Gustafsson08}{Gustafsson}, B., {Edvardsson}, B., {Eriksson}, K., et al., 2008, A\&A, 486, 951G
\bibitem[{Holtzman} et al.(2018)]{Holtzman18}{{Holtzman}, Jon A.,  {Hasselquist}, Sten, {Shetrone}, Matthew } et al., 2018, AJ, 156, 125H
\bibitem[J{\"o}nsson et al.(2018)]{Jonsson18}{{J{\"o}nsson}, Henrik, {Allende Prieto}, Carlos,
         {Holtzman}, Jon A., et al. 2018, AJ, 156, 126J}
\bibitem[Koleva et al.(2008)]{Koleva08}{Koleva}, M., {Prugniel}, P.,  {Ocvirk}, P., {Le Borgne}, D.,
	{Soubiran}, C., 2008, MNRAS, 385, 1998K
\bibitem[Kurucz(1979)]{Kurucz79}{Kurucz}, R.~L., 1979, ApJS, 40, 1K
\bibitem[Lan{\c{c}}on \& Wood(2000)]{Lancon2000}Lan{\c{c}}on, A., Wood, P. R., 2000, A\&AS, 146, 217
\bibitem[Law et al.(2016)]{Law16}Law, D. R., Cherinka, B., Yan, R., et al. 2016, AJ, 152, 83
\bibitem[Le Borgne et al.(2003)]{stelibref}{Le Borgne}, J.-F., {Bruzual}, G., {Pell{\'o}}, R., et al. 2003, A\&A, 402, 433L
\bibitem[Lejeune et al.(1998)]{Lejeune98}{Lejeune}, T., {Cuisinier}, F., {Buser}, R.,1998, A\&AS, 130, 65L
\bibitem[Leitherer et al.(2010)]{Leitherer10}{Leitherer}, Claus, {Ortiz Ot{\'a}lvaro}, Paula A. et al., 2010, ApJS, 189, 309L
\bibitem[Majewski et al.(2017)]{Majewski17}{Majewski}, Steven R.,  {Schiavon}, Ricardo P., {Frinchaboy}, Peter M., 2017, AJ, 154, 94M 
\bibitem[Maraston (2005)]{Maraston05}{Maraston}, Claudia, 2005, MNRAS, 362, 799M
\bibitem[Maraston \& {Str{\"o}mb{\"a}ck}(2011)]{Maraston11}{Maraston}, C., {Str{\"o}mb{\"a}ck}, G., 2011, MNRAS, 418, 2785M
\bibitem[Maraston et al.(2020 in press)]{Maraston20} Maraston, C., Hill, L., Thomas, D., et al.,  2020, MNRAS.tmp.1662M, (2019arXiv191105748M)
\bibitem[Martins et al.(2005)]{Martins05} Martins, L. P.,  {Gonz{\'a}lez Delgado}, R. M., Leitherer, C., {Cervi{\~n}o}, M., \& Hauschildt, P.,  2005, MNRAS, 358, 49M
\bibitem[Munari et al.(2005)]{Munari05}Munari, U., Sordo, R., Castelli, F., Zwitter, T. 2005, A\&A, 442,1127M
\bibitem[Neves et al.(2013)]{Neves13}{Neves}, V., {Bonfils}, X., {Santos}, N. C., {Delfosse}, X., {Forveille}, T., {Allard}, F., {Udry}, S.,2013, A\&A, 551A, 36N
\bibitem[Pickles(1985)]{Pickles85}{Pickles}, A. J., 1985, ApJS, 59, 33P
\bibitem[Pickles(1998)]{Pickles98}{Pickles}, A. J., 1998, PASP, 110, 863P
\bibitem[Prugniel \& Soubiran(2001)]{elodie}Prugniel, Ph., Soubiran, C. 2001, A\&A, 369, 1048P
\bibitem[Prugniel et al.(2007)]{Prugniel07}{Prugniel}, P.,  {Koleva}, M., {Ocvirk}, P.,  {Le Borgne}, D.,  {Soubiran}, C., 2007, IAUS, 241, 68P
\bibitem[Prugniel et al.(2011)]{Prugniel11}{Prugniel}, P., {Vauglin}, I., {Koleva}, M., 2011, A\&A, 531A, 165P
\bibitem[Rayner et al.(2009)]{Rayner09}Rayner, J. T., Cushing, M. C., Vacca, W. D. 2009, ApJS, 185, 289
\bibitem[S{\'a}nchez-Bl{\'a}zquez et al.(2006)]{milesref}{S{\'a}nchez-Bl{\'a}zquez}, P., {Peletier}, R. F.,  {Jim{\'e}nez-Vicente}, J., et al. 2006, MNRAS, 371, 703S	
\bibitem[Silva \&Cornell(1992)]{Silva92}Silva, D. R., Cornell, M. E., 1992, ApJS, 81, 865, doi: 10.1086/191706
\bibitem[Sharma et al.(2016)]{Sharma16}{Sharma}, K., {Prugniel}, P., {Singh}, H-P., 2016, A\&A, 585A, 64S
\bibitem[Valdes et al.(2004)]{Valdes04}Valdes, F., Gupta, R., Rose, J. A., Singh, H. P., Bell, D. J., 2004, ApJS, 152, 251, doi: 10.1086/386343
\bibitem[{Vazdekis} et al.(2010)]{Vazdekis10}{Vazdekis}, A., {S{\'a}nchez-Bl{\'a}zquez}, P., {Falc{\'o}n-Barroso}, J., et al.,2010, MNRAS, 404, 1639V 
\bibitem[Villaume et al.(2017)]{Villaume17}Villaume, A., Conroy, C., Johnson, B., et al. 2017, ApJS, 230, 23
\bibitem[Worthey et al.(1994)]{Worthey94}Worthey, G., Faber, S. M., Gonzalez, J. J., Burstein, D., 1994, ApJS, 94, 687
\bibitem[Wu et al.(2011)]{Wu11}{Wu}, Y., {Singh}, H. P., {Prugniel}, P., {Gupta}, R., {Koleva}, M., 2011, A\&A, 525A, 71W
\bibitem[Yan et al.(2016)]{Yan16}Yan, R., Tremonti, C., Bershady, M. A., et al. 2016, AJ, 151, 8Y
\bibitem[Yan et al.(2019)]{Yan19}Yan, R., Chen, Y.-P., Lazarz, D., et al., 2019, ApJ, .883, 175Y 
\bibitem[Yanny et al.(2009)]{yanny09}Yanny, B., Rockosi, C., Newberg, H. J., et al. 2009, AJ, 137, 4377, doi: 10.1088/0004-6256/137/5/4377
\bibitem[Zhao et al.(2012)]{zhao12}Zhao, G., Zhao, Y.-H., Chu, Y.-Q., Jing, Y.-P., Deng, L.-C. 2012, Research in Astronomy and Astrophysics, 12, 723, doi: 10.1088/1674-4527/12/7/002
\bibitem[Zwitter et al.(2004)]{Zwitter04}Zwitter, T., Castelli, F., Munari, U. 2004, A\&A, 417, 1055


\end{thebibliography}
\end{document}